\documentstyle[12pt,epsfig,axodraw]{article}

\renewcommand{\thefootnote}{\fnsymbol{footnote}}

\newcommand{\ra}{\rightarrow}
\newcommand{\ee}{e^+e^-}
\newcommand{\tb}{\tan\beta}
\newcommand{\s}{\\ \vspace*{-3mm} }
\newcommand{\nn}{\noindent}

\begin{document}

\newpage
\setcounter{page}{0}

\begin{titlepage}
\begin{flushright}
\hfill{YUMS 98--10}\\
\hfill{SNUTP 98--52}\\
\hfill{DESY 98--77}\\
\hfill{PM 98--14}\\
\hfill{IFT/9/98}\\

\hfill{June 1998}
\end{flushright}
\vspace*{1.0cm}

\begin{center}
{\large \bf Chargino Pair Production in $e^+e^-$ Collisions}
\end{center}
\vskip 1.cm
\begin{center}
{\sc S.Y. Choi$^1$, A. Djouadi$^2$, H. Dreiner$^3$}, 

\vspace*{0.4cm}

{\sc J. Kalinowski$^4$} and {\sc P.M. Zerwas$^5$} 

\vskip 0.8cm

\begin{small} 
$^1$ Department of Physics, Yonsei University, Seoul 120--749, Korea 
\vskip 0.2cm
$^2$ Phys. Math. et Th\'eorique, Universit\'{e} Montpellier II, 
F--34095 Montpellier, France 
\vskip 0.2cm
$^3$ Rutherford Laboratory, Chilton, Didcot OX11 OQX, U. Kingdom
\vskip 0.2cm
$^4$ Inst. Theor. Physics, Warsaw University, PL--00681 Warsaw, Poland 
\vskip 0.2cm
$^5$  DESY, Deutsches Elektronen-Synchrotron, D--22603 Hamburg, Germany 
\end{small}
\end{center}

\vskip 2cm

\setcounter{footnote}{0}
\begin{abstract}
Charginos $\chi_1^\pm$ are expected to be the lightest observable 
supersymmetric particles in many  supersymmetric models.
We present a procedure which will allow to determine the chargino 
mixing angles and, subsequently,  the fundamental SUSY parameters
$M_2, \mu$ and $\tb$ by measurements of the total cross section and 
the spin  correlations in $\ee$ annihilation to $\chi_1^+ \chi_1^-$ 
chargino pairs.
\end{abstract}
\end{titlepage}

\newpage
\renewcommand{\thefootnote}{\alph{footnote}}

\subsection*{1. Introduction}

In supersymmetric theories, the spin--1/2 partners of the $W$ boson and 
charged Higgs boson, $\tilde{W}^\pm$ and $\tilde{H}^\pm$, mix to form 
chargino mass eigenstates $\tilde{\chi}^\pm_{1,2}$. The mass eigenvalues 
$m_{\tilde{\chi}_{1,2}^\pm}$ and the mixing angles $\phi_L,\phi_R$
are determined by the elements of the chargino mass matrix in the
$(\tilde{W}^-,\tilde{H}^-)$ basis \cite{R1} 
\begin{eqnarray}
{\cal M}_C=\left(\begin{array}{cc}
                M_2                &      \sqrt{2}m_W\cos\beta  \\
             \sqrt{2}m_W\sin\beta  &             \mu   
                  \end{array}\right)
\label{eq:mass matrix}
\end{eqnarray}
which is built up by the fundamental supersymmetric parameters\footnote{The
chargino/neutralino sector is assumed to be CP--invariant in the following
analysis. The consequences of CP non--invariance will be discussed briefly
in the
Appendix.}: the gaugino
mass $M_2$, the Higgs mass parameter $\mu$, and the ratio $\tb=v_2/v_1$
of the vacuum expectation values of the two neutral Higgs fields which
break the electroweak symmetry. Once charginos will have been discovered, 
the experimental analysis of their properties, production and decay 
mechanisms, will therefore reveal the basic structure of the underlying 
low--energy supersymmetric theory. \s

Charginos are produced in $e^+e^-$ collisions, either in diagonal or in mixed 
pairs \cite{R2}. In the present note, we will focus on the 
diagonal pair production 
of the lightest chargino $\tilde{\chi}_1^\pm$ in $e^+e^-$ collisions, 
\begin{eqnarray*}
e^+e^- \ \rightarrow \ \tilde{\chi}^+_1 \ \tilde{\chi}^-_1 
\end{eqnarray*}
The second chargino $\tilde{\chi}_2^\pm$ is generally expected to be 
significantly heavier than the first state. At LEP2 \cite{R3}, and 
potentially even in the first phase of  $e^+e^-$ linear colliders (see
e.g. Ref.~\cite{R4}), the chargino $\tilde{\chi}_1^\pm$ may  be, 
for some time,  the
only chargino state  that can be studied experimentally 
in  detail. \s 

Even in this situation the underlying fundamental parameters can be extracted 
from the mass $m_{\tilde{\chi}^\pm_1}$, the total production cross section, 
and the measurement of the polarization with 
which the charginos are produced. \s
 
The $\tilde{\chi}$ polarization vectors and
$\tilde{\chi}$--$\tilde{\chi}$ spin--spin correlation tensors can be
determined from the decay distributions of the charginos. Beam
polarization is helpful but not necessarily required. We will assume
that the charginos decay into the lightest neutralino
$\tilde{\chi}^0_1$, which is taken to be stable, and a pair of quarks
and antiquarks or leptons and neutrinos:
$\tilde{\chi}^\pm_1\rightarrow \tilde{\chi}^0_1+f\bar{f}'$. It is very
important to note, however, that no detailed information on the decay
dynamics, nor on the structure of the neutralino, is needed to carry
out the spin analysis \cite{R4A}. Thus the analysis of the chargino
properties is separated from the neutralino sector. Since two neutral
particles $\tilde{\chi}^0_1$ escape undetected, it is not possible to
reconstruct the events unambiguously. The partial information on the
chargino polarizations is nevertheless sufficient to extract the
fundamental supersymmetric parameters up to at most a two--fold
discrete ambiguity. In contrast to earlier analyses \cite{R6, R6A}, we
will not elaborate on global chargino/neutralino fits but rather
attempt to explore the event characteristics to isolate the chargino
sector. \s

The analysis will be based strictly on low--energy supersymmetry. Once
these parameters will have been extracted experimentally, they may be
confronted with the relations as predicted in Grand Unified Theories
for instance. The paper will be divided into four parts. In Section~2
we briefly recapitulate the elements of the mixing formalism for the
sake of convenience. In Section~3 the cross sections for chargino
production and the chargino polarization vectors are given. The
analysis power for measuring the chargino polarization vectors and
spin correlations is exemplified for appropriate decay modes in
Section~4. In Section~5 we describe a set of observables which can be
used in measurements of angular correlations to extract the
fundamental supersymmetric parameters in a model--independent way.
Conclusions are given in Section 6. In an appendix, we discuss the
impact of potential CP non--invariance in the chargino/neutralino
sector on the present analysis.

\subsection*{2. Mixing Formalism}
\label{sec:mixing}
 
Since the chargino mass matrix ${\cal M}_C$ is not 
symmetric, two different matrices acting on the left-- and right--chiral 
$(\tilde{W},\tilde{H})$ states are needed to diagonalize the matrix. 
The lightest of the two eigenvalues is given by \cite{R1}
\begin{eqnarray}
m^2_{\tilde{\chi}^\pm_1}
  =\frac{1}{2}\left[M^2_2+\mu^2+2m^2_W
    -\sqrt{(M^2_2+\mu^2+2m^2_W)^2-4(M_2\mu-m^2_W\sin 2\beta)^2} \, \right]
\end{eqnarray}
The left-- and right--chiral components of the mass eigenstate 
$\tilde{\chi}^-_1$ are related to the wino and higgsino components
in the following way,  
\begin{eqnarray}
&&\tilde{\chi}^-_{1L}=\tilde{W}^-_L\cos\phi_L
                     +\tilde{H}^-_{1L}\sin\phi_L \nonumber\\
&&\tilde{\chi}^-_{1R}=\tilde{W}^-_R\cos\phi_R
                     +\tilde{H}^-_{2R}\sin\phi_R 
\end{eqnarray}
with the rotation angles
\begin{eqnarray}
&&\cos 2\phi_L=-\frac{M_2^2-\mu^2-2m^2_W\cos 2\beta}{
             \sqrt{(M^2_2+\mu^2+2m^2_W)^2-4(M_2\mu-m^2_W\sin 2\beta)^2}}
\nonumber\\ 
&&\sin 2\phi_L=-\frac{2\sqrt{2}m_W(M_2\cos\beta+\mu\sin\beta)}{
             \sqrt{(M^2_2+\mu^2+2m^2_W)^2-4(M_2\mu-m^2_W\sin 2\beta)^2}}
\nonumber
\end{eqnarray}
and
\begin{eqnarray}
&&\cos 2\phi_R=-\frac{M_2^2-\mu^2+2m^2_W\cos 2\beta}{
             \sqrt{(M^2_2+\mu^2+2m^2_W)^2-4(M_2\mu-m^2_W\sin 2\beta)^2}}
\nonumber\\ 
&&\sin 2\phi_R=-\frac{2\sqrt{2}m_W(M_2\sin\beta+\mu\cos\beta)}{
             \sqrt{(M^2_2+\mu^2+2m^2_W)^2-4(M_2\mu-m^2_W\sin 2\beta)^2}} 
\end{eqnarray}
As usual, we take $\tan\beta$ positive, $M_2$ positive and $\mu$ of either
sign.

The three fundamental supersymmetric parameters $M_2$, $\mu$ and
$\tan\beta$ can be extracted from the three chargino
$\tilde{\chi}^\pm_1$ parameters: the mass $m_{\tilde{\chi}^\pm_1}$ and
the two mixing angles $\phi_L$ and $\phi_R$ of the left-- and
right--chiral components of the wave function. These mixing angles are
physical observables and they can be measured in the process
$e^+e^-\rightarrow\tilde{\chi}^+_1\tilde{\chi}^-_1$ if the
polarization states of the charginos are analyzed. \s

The two angles $\phi_L$ and $\phi_R$ define the couplings of the 
chargino--chargino--$Z$ vertices and the electron--sneutrino--chargino vertex:
\begin{eqnarray}
&&\langle\tilde{\chi}^-_{1L}|Z|\tilde{\chi}^-_{1L}\rangle 
  = -\frac{e}{s_W c_W} \left[s_W^2 - \frac{3}{4}-\frac{1}{4}
\cos 2\phi_L\right] \nonumber\\
&&\langle\tilde{\chi}^-_{1R}|Z|\tilde{\chi}^-_{1R}\rangle 
  = -\frac{e}{s_W c_W} \left[s_W^2-\frac{3}{4}-\frac{1}{4}\cos 
2\phi_R\right] \nonumber\\
&&\langle\tilde{\chi}^-_{1R}|\tilde{\nu}|e^-_L\rangle 
  =-\frac{e}{s_W}\cos\phi_R
\label{eq:vertex}
\end{eqnarray}
where $s_W^2 =1-c_W^2 \equiv \sin^2\theta_W$. The coupling to the
higgsino component, being proportional to the electron mass, has been
neglected in the sneutrino vertex; the sneutrino couples only to
left--handed electrons. Since the photon--chargino vertex is diagonal,
it does not depend on the mixing angles:
\begin{eqnarray}
\langle\tilde{\chi}^-_{1L,R}|\gamma|\tilde{\chi}^-_{1L,R}\rangle =e 
\end{eqnarray}
The parameter $e$ is the electromagnetic coupling which will   
be defined at an effective scale 
which is identified with the c.m. energy $\sqrt{s}$.

\subsection*{3. The Production of Polarized Charginos}
\label{sec:production}

The process $e^+e^-\rightarrow\tilde{\chi}^+_1\tilde{\chi}^-_1$ is generated
by the three mechanisms shown in Fig.~1: $s$--channel $\gamma$ and $Z$
exchanges, and $t$--channel $\tilde{\nu}$ exchange. The transition matrix
element, after a Fierz transformation of the $\tilde{\nu}$--exchange
amplitude,
\begin{eqnarray}
T\left(e^+e^-\rightarrow\tilde{\chi}^+_1\tilde{\chi}^-_1\right)
 = \frac{e^2}{s}Q_{\alpha\beta}
   \left[\bar{v}(e^+)  \gamma_\mu P_\alpha  u(e^-)\right]
   \left[\bar{u}(\tilde{\chi}^-_1) \gamma^\mu P_\beta 
               v(\tilde{\chi}^+_1)\right]
\label{eq:production amplitude}
\end{eqnarray}
can be expressed in terms of four 
bilinear charges, classified according to the chiralities $\alpha,
\beta=L,R$ of the associated lepton and chargino currents
\begin{eqnarray}
Q_{LL}&=&1+ \frac{D_Z}{s_W^2 c_W^2}(s_W^2 -\frac{1}{2}) 
         \left(s_W^2 -\frac{3}{4}-\frac{1}{4}\cos 2\phi_L\right) 
         \nonumber\\ 
Q_{LR}&=&1+ \frac{D_Z}{s_W^2 c_W^2} (s_W^2 -\frac{1}{2}) 
         \left(s_W^2-\frac{3}{4}-\frac{1}{4}\cos 2\phi_R\right) 
        + \frac{D_{\tilde{\nu}}}{4s_W^2} (1+\cos 2\phi_R) \nonumber\\
Q_{RL}&=&1+\frac{D_Z}{c_W^2} \left(s_W^2 -\frac{3}{4}-\frac{1}{4}\cos 
          2\phi_L\right) \nonumber\\
Q_{RR}&=&1+ \frac{D_Z}{c_W^2}  \left(s_W^2 -\frac{3}{4}-\frac{1}{4}\cos 
         2\phi_R\right)
\end{eqnarray}
The first index in $Q_{\alpha \beta}$ refers to the chirality of the $e^\pm$ 
current, the second index to the chirality of the $\tilde{\chi}_1^\pm$ 
current. The $\tilde{\nu}$ 
exchange affects only the $LR$ chirality charge while all other amplitudes 
are built up by $\gamma$ and $Z$ exchanges. $D_{\tilde{\nu}}$ denotes the
sneutrino propagator $D_{\tilde{\nu}} = s/(t- m_{\tilde{\nu}}^2)$, and 
$D_Z$ the $Z$ propagator $D_Z=s/(s-m^2_Z+im_Z\Gamma_Z)$; the non--zero width 
can in general be neglected for the energies considered in the present 
analysis so that the charges are real. \s 

For the sake of convenience we also introduce the quartic charges \cite{R7}
\begin{eqnarray}
&&    Q_1=\frac{1}{4}\left[|Q_{RR}|^2+|Q_{LL}|^2
                          +|Q_{RL}|^2+|Q_{LR}|^2\right] \nonumber\\
&&    Q_2=\frac{1}{2}{\rm Re}\left[Q_{RR}Q^*_{RL}
                                  +Q_{LL}Q^*_{LR}\right] \nonumber\\
&&    Q_3=\frac{1}{4}\left[|Q_{RR}|^2+|Q_{LL}|^2
                          -|Q_{RL}|^2-|Q_{LR}|^2\right] 
\end{eqnarray}
and
\begin{eqnarray}
&&   Q'_1=\frac{1}{4}\left[|Q_{RR}|^2+|Q_{RL}|^2
                          -|Q_{LR}|^2-|Q_{LL}|^2\right]\nonumber\\
&&   Q'_2=\frac{1}{2}{\rm Re}\left[Q_{RR}Q^*_{RL}
                                  -Q_{LL}Q^*_{LR}\right]\nonumber\\
&&   Q'_3=\frac{1}{4}\left[|Q_{RR}|^2+|Q_{LR}|^2
                          -|Q_{RL}|^2-|Q_{LL}|^2\right]
\end{eqnarray}
The measurement of the quartic charges $Q_1$ to $Q'_3$ will allow us to
extract the two terms $\cos 2\phi_L$ and $\cos 2\phi_R$ unambiguously.
The corresponding quantities $\sin 2\phi_L$ and $\sin 2\phi_R$ are 
determined up to a sign ambiguity.

\vspace*{1cm} 
\begin{center}
\begin{picture}(330,100)(0,0)
\Text(15,85)[]{$e^-$}
\ArrowLine(10,75)(35,50)
\ArrowLine(35,50)(10,25)
\Text(15,15)[]{$e^+$}
\Photon(35,50)(75,50){4}{8}
\Text(55,37)[]{$\gamma$}
\ArrowLine(75,50)(100,75)
\Photon(75,50)(100,75){3}{7}
\Text(95,85)[]{$\tilde{\chi}^-_1$}
\ArrowLine(100,25)(75,50)
\Photon(100,25)(75,50){3}{7}
\Text(95,15)[]{$\tilde{\chi}^+_1$}
\Text(125,85)[]{$e^-$}
\ArrowLine(120,75)(145,50)
\Text(125,15)[]{$e^+$}
\ArrowLine(145,50)(120,25)
\Photon(145,50)(185,50){4}{8}
\Text(165,37)[]{$Z$}
\ArrowLine(185,50)(210,75)
\Photon(185,50)(210,75){3}{7}
\Text(207,85)[]{$\tilde{\chi}^-_1$}
\ArrowLine(210,25)(185,50)
\Photon(210,25)(185,50){3}{7}
\Text(207,15)[]{$\tilde{\chi}^+_1$}
\Text(235,85)[]{$e^-$}
\ArrowLine(230,75)(275,75)
\Text(235,15)[]{$e^+$}
\ArrowLine(275,25)(230,25)
\Line(274,75)(274,25)
\Line(276,75)(276,25)
\Text(285,50)[]{$\tilde{\nu}$}
\ArrowLine(275,75)(320,75)
\Photon(275,75)(320,75){3}{7}
\Text(318,85)[]{$\tilde{\chi}^-_1$}
\ArrowLine(320,25)(275,25)
\Photon(320,25)(275,25){3}{7}
\Text(318,15)[]{$\tilde{\chi}^+_1$}
\end{picture}\\
\end{center}
\smallskip

Figure~1: {\it The three mechanisms contributing to the production of 
            diagonal chargino  \\ \hspace*{0.41cm} pairs 
            $\tilde{\chi}^+_1 \tilde{\chi}^-_1$ in $\ee$ annihilation.}
\bigskip 

Defining the $\tilde{\chi}^-_1$ production angle with respect to the
electron flight--direction by $\Theta$, the helicity amplitudes can be
determined from eq.~(\ref{eq:production amplitude}). While electron
and positron helicities are opposite to each other in all amplitudes,
the $\tilde{\chi}^-_1$ and $\tilde{\chi}^+_1$ helicities are in
general not correlated due to the non--zero masses of the particles;
amplitudes with equal $\tilde{\chi}^\pm_1$ helicities vanish only
$\propto m_{\tilde{\chi}^\pm_1} /\sqrt{s}$ for asymptotic energies.
Denoting the electron helicity by the first index, the
$\tilde{\chi}^-_1$ and $\tilde{\chi}^+_1$ helicities by the remaining
two indices, the helicity amplitudes $T(\sigma;\lambda,
\bar{\lambda})=2\pi\alpha\langle\sigma;\lambda\bar{\lambda}\rangle$
are given as follows \cite{R8},
\begin{eqnarray}
&& \langle +;++\rangle 
   =-\sqrt{1-\beta^2}\left[Q_{RR}+Q_{RL}\right]\sin\Theta \nonumber\\
&& \langle +;+-\rangle 
   =-\left[(1+\beta)Q_{RR}+(1-\beta)Q_{RL}\right](1+\cos\Theta) \nonumber\\
&& \langle +;-+\rangle 
   =+\left[(1-\beta)Q_{RR}+(1+\beta)Q_{RL}\right](1-\cos\Theta) \nonumber\\
&& \langle +;--\rangle 
   =+\sqrt{1-\beta^2}\left[Q_{RR}+Q_{RL}\right]\sin\Theta 
\end{eqnarray}
and
\begin{eqnarray}
&& \langle -;++\rangle 
   =-\sqrt{1-\beta^2}\left[Q_{LR}+Q_{LL}\right]\sin\Theta \nonumber\\
&& \langle -;+-\rangle 
   =+\left[(1+\beta)Q_{LR}+(1-\beta)Q_{LL}\right](1-\cos\Theta) \nonumber\\
&& \langle -;-+\rangle 
   =-\left[(1-\beta)Q_{LR}+(1+\beta)Q_{LL}\right](1+\cos\Theta) \nonumber\\
&& \langle -;--\rangle 
   =+\sqrt{1-\beta^2}\left[Q_{LR}+Q_{LL}\right]\sin\Theta 
\label{eq:helicity amplitude}
\end{eqnarray}
where $\beta=\sqrt{1-4m^2_{\tilde{\chi}^\pm_1}/s}$ is the $\tilde{\chi}^\pm_1$
velocity in the c.m.~frame. From these amplitudes the production cross 
section, the $\tilde{\chi}^-_1$ and $\tilde{\chi}^+_1$ polarization vectors 
and the $\tilde{\chi}$--$\tilde{\chi}$ spin--spin correlation tensors can be 
determined. \s

The final state probability may be expanded in terms of the
unpolarized cross section, the polarization vectors of
$\tilde{\chi}^-_1$ and $\tilde{\chi}^+_1$, and the spin--spin
correlation tensor. Defining the $\hat{z}$ axes by the
$\tilde{\chi}^\pm$ momenta, the $\hat{x}$ axes in the production plane
(rotated counter-clockwise by $90^o$ from the $\tilde{\chi}^-$ flight
direction), and $\hat{y} =\hat{z}\times\hat{x}$ in the rest frames of
the charginos, cross section and spin--density matrices may be written
as \cite{R9}:
\begin{eqnarray}
&&\frac{{\rm d}\sigma}{{\rm d}\cos\Theta}
     (\lambda\lambda';\bar{\lambda}\bar{\lambda}')
 =\frac{\pi\alpha^2}{32 s} \beta 
   \sum_{\sigma=\pm}\langle\sigma;\lambda\bar{\lambda}\rangle
              \langle\sigma;\lambda'\bar{\lambda}'\rangle^* \\
&& { }\hskip 0.8cm =\frac{{\rm d}\sigma}{{\rm d}\cos\Theta}\frac{1}{4}
   \bigg[(I)_{\lambda'\lambda}(I)_{\bar{\lambda}\bar{\lambda}'}
  +{\cal P}_i (\tau^i)_{\lambda'\lambda}(I)_{\bar{\lambda}\bar{\lambda}'}
  +\bar{\cal P}_i (I)_{\lambda'\lambda}(\tau^i)_{\bar{\lambda}\bar{\lambda}'}
  +{\cal Q}_{ij}(\tau^i)_{\lambda'\lambda}(\tau^j)_{\bar{\lambda}
   \bar{\lambda}'}
   \bigg] \nonumber 
\end{eqnarray}
$\lambda(\lambda')$ and $\bar{\lambda}(\bar{\lambda}')$ 
are twice the helicities, $\pm 1$, of the $\tilde{\chi}^-_1$ 
and $\tilde{\chi}^+_1$ particles in the final state. The $\tau^i$ are the
Pauli matrices with respect to the reference frame introduced above. \s

Alternatively, the polarization vectors and the spin--spin correlation matrix
may be defined in the following covariant way. Denoting the $\tilde{\chi}^-_1$
spin--quantization axis by $n_\mu$, the $\tilde{\chi}^+_1$ axis by $\bar{n}
_\mu$, the cross section for $\ee \ra \tilde{\chi}^-_1(n)\tilde{\chi}^+_1
(\bar{n})$ may be written \cite{R9A} 
\begin{eqnarray}
\frac{{\rm d}\sigma}{{\rm d}\cos\Theta}(n,\bar{n})
 =\frac{{\rm d}\sigma}{{\rm d}\cos\Theta}\frac{1}{4}\bigg[
  1-{\cal P}^\mu n_\mu -\bar{\cal P}^\mu\bar{n}_\mu
 +{\cal Q}^{\mu\nu} n_\mu\bar{n}_\nu\bigg]
\end{eqnarray}
The two representations are related through the identities
\begin{eqnarray}
 {\cal P}_i= - {\cal P}_\mu\eta^\mu_i \ \ {\rm and} \ \ 
\bar{\cal P}_i= - \bar{\cal P}_\mu\bar{\eta}^\mu_i \nonumber\\
 {\cal Q}_{ij}={\cal Q}_{\mu\nu}\eta^\mu_i\bar{\eta}^\nu_j \hspace*{1cm}
\label{eq:polarization vector1}
\end{eqnarray}
with $\eta_i(\bar{\eta}_i)$ being  the 
three unit vectors in the particle (antiparticle) rest frame 
Lorentz--boosted to
the laboratory frame. 

\subsubsection*{3.1 The production cross section}

The unpolarized differential cross section is given by the average/sum over 
the helicities:
\begin{eqnarray}
\frac{{\rm d}\sigma}{{\rm d}\cos\Theta}
      (e^+e^-\rightarrow\tilde{\chi}^+_1\tilde{\chi}^-_1)
 =\frac{\pi\alpha^2}{32 s} \beta \, 
  \sum_{\sigma\lambda\bar{\lambda}}\,
  |\langle\sigma;\lambda\bar{\lambda}\rangle|^2
\end{eqnarray}
Carrying out the sum, one finds the following expression for the cross 
section in terms of the quartic charges:
\begin{eqnarray}
\frac{{\rm d}\sigma}{{\rm d}\cos\Theta}
      (e^+e^-\rightarrow\tilde{\chi}^+_1\tilde{\chi}^-_1)
 =\frac{\pi\alpha^2}{2 s} \beta 
  \left\{(1+\beta^2\cos^2\Theta)Q_1+(1-\beta^2)Q_2
         +2\beta\cos\Theta Q_3\right\}
\label{eq:cross section}
\end{eqnarray}
If the production angle could be measured unambiguously 
on an event--by--event basis, the quartic charges could be extracted directly 
from the angular dependence of the cross section. \s

The total production cross section is shown in Fig.~2 as a function of (a) 
the c.m.~energy for a fixed sneutrino mass, and (b) the sneutrino mass at 
the c.m. energy of 200 GeV for a representative set of parameters.
The parameters are chosen in the higgsino region $M_2 \gg |\mu|$, the gaugino 
region $M_2 \ll |\mu|$ and in the mixed region $M_2 \sim |\mu|$ for 
$\tan\beta=2$ as
\begin{eqnarray}
\begin{array}{ll}
{\rm gaugino\ \ region} \makebox[1mm]{}:
 & (M_2,\mu)=(81\ \ {\rm GeV}, -215\ \ {\rm GeV})\\
{\rm higgsino\ \  region}: & (M_2,\mu)=(215\ \ {\rm GeV}, -81\ \ {\rm GeV})\\
{\rm mixed\ \ region} \makebox[4mm]{}: & (M_2,\mu)=(92\ \ {\rm GeV}, 
  -93\ \ {\rm GeV})
\label{eq:parameter}
\end{array}
\end{eqnarray}
for which the light chargino mass $m_{\tilde{\chi}^\pm_1}$ is approximately 
$95$ GeV. The sharp rise of the production cross section in Fig.~2a  
 allows to measure  the chargino mass $m_{\tilde{\chi}^\pm_1}$ very 
precisely. In
Fig.~2b it is shown that the $\tilde{\nu}$--exchange diagram, as well-known,
leads to a strong destructive interference for the gaugino and mixed regions, 
while the dependence of the cross section on $m_{\tilde{\nu}}$ decreases as 
the higgsino component of the chargino increases. 
Prior or simultaneous determination of $m_{\tilde{\nu}}$ is therefore 
necessary to determine the other SUSY parameters. \s

Fig.~3 exhibits the angular distribution as a function of the 
scattering angle for the same parameters as in eq.~(\ref{eq:parameter})
at a c.m. energy of (a) 200 GeV and (b) 500 GeV. 
The angular distribution depends strongly on the $(M_2, \mu)$ parameter values.
The peak in the forward region for the gaugino and mixed points is due 
to the $t$-channel sneutrino exchange; the distribution is almost 
forward-backward symmetric in the higgsino scenario. 

\subsubsection*{3.2 The chargino polarization vectors}
\label{subsec:polarization vector}

The polarization vector $\vec{\cal P}=({\cal P}_T,{\cal P}_N,
{\cal P}_L)$ is defined in the rest frame of the particle $\tilde{\chi}^-_1$. 
${\cal P}_L$ denotes the component parallel to the $\tilde{\chi}^-_1$ flight 
direction in the c.m. frame, ${\cal P}_T$ the transverse component in the 
production plane, and ${\cal P}_N$ the  component normal to the production 
plane. These three components can be expressed by helicity amplitudes in the 
following way:
\begin{eqnarray}
&& {\cal P}_L   = \frac{1}{4}\sum_{\sigma=\pm} \left\{
          |\langle\sigma;++\rangle|^2+|\langle\sigma;+-\rangle|^2 
         -|\langle\sigma;-+\rangle|^2-|\langle\sigma;--\rangle|^2\right\}
/{\cal N}            \nonumber\\
&& {\cal P}_T =\frac{1}{2}{\rm Re}\left\{\sum_{\sigma=\pm}\left[
           \langle\sigma;++\rangle\langle\sigma;-+\rangle^* 
          +\langle\sigma;--\rangle\langle\sigma;+-\rangle^*\right]\right\}
/{\cal N}            \nonumber\\
&& {\cal P}_N =\frac{1}{2}{\rm Im}\left\{\sum_{\sigma=\pm}\ \left[
           \langle\sigma;--\rangle\langle\sigma;+-\rangle^* 
          -\langle\sigma;++\rangle\langle\sigma;-+\rangle^*\right]\right\}
/{\cal N}            
\label{eq:polarization vector}
\end{eqnarray}
with the normalization
\begin{eqnarray}
{\cal N} =\frac{1}{4}
\sum \left[|\langle\sigma;++\rangle|^2+|\langle\sigma;+-\rangle|^2 
         +|\langle\sigma;-+\rangle|^2+|\langle\sigma;--\rangle|^2\right]
\end{eqnarray}
The corresponding polarization 4--vectors can readily be expressed in terms of 
the quartic charges,
\begin{eqnarray}
{\cal P}_\mu     =\frac{8m_{\tilde{\chi}^-_1}}{s} \left\{
            (\bar{l}-l)_\mu[{Q'}_1+{Q'}_2+\beta\cos\Theta{Q'}_3]
           +(l+\bar{l})_\mu[{Q'}_3+\beta\cos\Theta({Q'}_1-{Q'}_2)]\right\}/
            {\cal N} \nonumber\\
\bar{\cal P}_\mu =\frac{8m_{\tilde{\chi}^+_1}}{s} \left\{
            (\bar{l}-l)_\mu[{Q'}_1+{Q'}_2+\beta\cos\Theta{Q'}_3]
           -(l+\bar{l})_\mu[{Q'}_3+\beta\cos\Theta({Q'}_1-{Q'}_2)]\right\}/
            {\cal N} 
\nonumber\\ 
\end{eqnarray}
with, correspondingly,
\begin{eqnarray}
{\cal N}=4\{(1+\beta^2\cos^2\Theta)Q_1+(1-\beta^2)Q_2+2\beta\cos\Theta Q_3\}
\end{eqnarray}
The vectors $l_\mu$ and $\bar{l}_\mu$ are the 4--momenta of the incoming 
electrons and positrons, respectively. \s

The normal component can only be generated by complex production amplitudes.
Non-zero phases are present in the fundamental supersymmetric parameters
if CP is broken in the supersymmetric interaction \cite{R1}. Also, the 
non--zero width of the $Z$ boson and loop corrections generate non--trivial 
phases; however, the width effect is negligible for high energies and the 
effects due to radiative corrections are small. Neglecting loops 
and the small $Z$--width, the normal $\tilde{\chi}^-_1$ and 
$\tilde{\chi}^+_1$ polarizations are zero  since
the $\tilde{\chi}_1 \tilde{\chi}_1 \gamma$ and 
$\tilde{\chi}_1 \tilde{\chi}_1 Z$ vertices 
are real even for non-zero phases in the chargino mass matrix, 
and the sneutrino--exchange amplitude is real too. The CP--violating phases 
will change the chargino mass and the mixing angles \cite{R11} but they do not 
induce complex charges in the production amplitudes of the light chargino 
pairs (see Appendix). \s

The longitudinal and transverse components of the $\tilde{\chi}_1^-$
polarization vector can easily be obtained from the helicity amplitudes
or from  the covariant representation: 
\begin{eqnarray}
&& {\cal P}_L    =4\left\{ (1+\beta^2)\cos\Theta{Q'}_1
                      +(1-\beta^2)\cos\Theta {Q'}_2
                      +(1+\cos^2\Theta)\beta {Q'}_3 \right\} / {\cal N}
\nonumber \\
&& {\cal P}_T  =-4\sqrt{1-\beta^2}\sin\Theta
                 \left\{ {Q'}_1+{Q'}_2+\beta\cos\Theta{Q'}_3
                 \right\}/{\cal N} 
\end{eqnarray}
where the normalization ${\cal N}$ is given in eq.~(22). 
The polarization vector depends on the quartic charges $Q'_1$ to $Q'_3$,
which are independent out of the charges $Q_1$ and $Q_3$.
Representative examples of their size are shown as a function of $\cos\Theta$
in Fig.~4;  the same parameters are adopted as for the cross 
section in Fig.~2.  The dependence of the longitudinal and 
transverse polarizations on the SUSY parameters 
is rather weak at $\sqrt{s}=200$ GeV; 
close to the production threshold, 
${\cal P}_L$ and ${\cal P}_R$ are given by the same combination
of quartic charges:
\begin{eqnarray}
{\cal P}_L \rightarrow  \frac{Q'_1+Q'_2}{Q_1+Q_2} \cos\Theta 
\makebox[6mm]{} \makebox{and} \makebox[6mm]{}
{\cal P}_T \rightarrow -\frac{Q'_1+Q'_2}{Q_1+Q_2} \sin\Theta
\end{eqnarray}
The sensitivity is stronger at $\sqrt{s}=500$ GeV where the gaugino scenario 
is clearly separated from the higgsino scenario.

\subsubsection*{3.3 Chargino spin--correlations}
\label{subsec:spin correlation}

The three quartic charges $Q_1$, $Q_2$ and $Q_3$ determine the
$\Theta$ dependence of the cross section. This would be sufficient for 
measuring the charges if the production angle $\Theta$ could be determined 
unambiguously on an event--by--event basis. However, this is not possible
due to the two LSP's which escape detection. Additional information
on these three quartic charges can however be obtained from the observation of 
spin--spin correlations. Since they are reflected in the angular correlations
between the $\tilde{\chi}^-_1$ and $\tilde{\chi}^+_1$ decay products, they are experimentally accessible directly. Moreover, any dependence on the 
specific parameters of the decay mechanisms can be eliminated as shown later
in detail. \s

The spin--spin correlation matrix ${\cal Q}_{ij}$ consists of nine independent 
elements. They can be derived from the $\eta_i \times \bar{\eta}_j$ 
projections of the covariant matrix ${\cal Q}_{\mu \nu}$:
\begin{eqnarray}
{\cal Q}_{\mu\nu}&=&\frac{4}{\cal N}\bigg\{
     g_{\mu\nu}(1-\cos^2\Theta)\beta^2 Q_2\nonumber\\
  &&-\frac{2}{s}l_\mu\bar{l}_\nu
     \left[(1+\beta^2+2\beta\cos\Theta)Q_2+(1-\beta^2)(Q_1-Q_3)\right]
     \nonumber\\
  &&-\frac{2}{s}l_\nu\bar{l}_\mu
     \left[(1+\beta^2-2\beta\cos\Theta)Q_2+(1-\beta^2)(Q_1+Q_3)\right]
     \bigg\}
\end{eqnarray}
Note that the spin-spin correlation matrix is built up again by the 
same quartic charges $Q_1$, $Q_2$ and $Q_3$ as the unpolarized cross section.

\subsection*{4. Chargino Decays and Correlations}
\label{sec:decay}

\subsubsection*{4.1 Chargino Decays}

The polarization and spin--spin correlations of the charginos can be
inferred from the angular distributions of the decay products.
Assuming the  neutralino $\tilde{\chi}^0_1$ to be the
lightest supersymmetric particle,  
several mechanisms contribute to the decay of the chargino
$\tilde{\chi}^-_1$:
\begin{eqnarray*}
\tilde{\chi}^-_1 \rightarrow\tilde{\chi}^0_1 (q_0) +f_1
(q) \bar{f}_2 (\bar{q}) \ \ \hskip 1cm
                             [f_i=l,\nu,q]
\end{eqnarray*}
\vspace*{-1.9cm}
\begin{center}
\begin{picture}(350,220)(0,0)
\Text(0,120)[]{$\tilde{\chi}^-_1$}
\ArrowLine(10,120)(40,120)
\ArrowLine(40,120)(70,150)
\Photon(40,120)(70,150){3}{7}
\Text(77,150)[]{$\tilde{\chi}^0_1$}
\Photon(40,120)(60,90){4}{8}
\Text(30,95)[]{$W^-$}
\ArrowLine(60,90)(90,90)
\Text(100,90)[]{$d$}
\ArrowLine(80,60)(60,90)
\Text(85,55)[]{$\bar{u}$}
\Text(125,120)[]{$\tilde{\chi}^-_1$}
\ArrowLine(135,120)(165,120)
\ArrowLine(165,120)(195,150)
\Text(200,150)[]{$d$}
\Line(185,90)(165,120)
\Line(188,90)(167,122)
\Text(165,95)[]{$\tilde{u}_L$}
\ArrowLine(185,90)(215,90)
\Photon(185,90)(215,90){3}{7}
\Text(225,90)[]{$\tilde{\chi}^0_1$}
\ArrowLine(205,60)(185,90)
\Text(210,55)[]{$\bar{u}$}
\Text(250,120)[]{$\tilde{\chi}^-_1$}
\ArrowLine(260,120)(290,120)
\ArrowLine(320,150)(290,120)
\Text(325,150)[]{$\bar{u}$}
\Line(290,120)(310,90)
\Line(292,122)(313,90)
\Text(290,95)[]{$\tilde{d}_L$}
\ArrowLine(310,90)(340,90)
\Photon(310,90)(340,90){3}{7}
\Text(348,90)[]{$\tilde{\chi}^0_1$}
\ArrowLine(310,90)(335,55)
\Text(335,55)[]{$d$}
\end{picture}\\
\end{center}
\vspace*{-0.6cm}
Figure~5: {\it Chargino decay mechanisms; the exchange of the charged Higgs 
               boson is \\ \hspace*{0.4cm} neglected.}
\bigskip

The corresponding diagrams are shown in Fig.~5 for the decay into quark
pairs. The exchange of  the charged Higgs boson [replacing the $W$
boson] can be  neglected since the couplings to the light SM leptons and 
quarks are 
very small. In this case, all the components of the decay matrix elements 
are of the left/right current$\times$current form which, after a simple 
Fierz transformation, may be written for quark final states as\footnote{If
$m_{\chi_1^\pm} > m_{\tilde \nu}$, the two--body decay of the
chargino into a sneutrino and a charged lepton (with the sneutrino
subsequently decaying into a neutrino and the lightest neutralino) will 
be the dominant mode \cite{R11A}. This case can be taken into account by
including the decay width of the sneutrino in the propagators.}:
\begin{eqnarray}
{\cal D}~\left[\tilde{\chi}^-_1\rightarrow\tilde{\chi}^0_1d\bar{u}\right]
 =\frac{e^2}{2 \sqrt{2}s_W^2}\bigg[\bar{u}(\tilde{\chi}^0_1)\gamma_\mu
                   [F_LP_L+F_RP_R] u(\tilde{\chi}_1)\bigg]
                       \bigg[\bar{u}(d)\gamma^\mu P_L v(\bar{u})\bigg]
\end{eqnarray}
with 
\begin{eqnarray}
F_L&=&\frac{2 N_{12} \cos\phi_L + \sqrt{2} N_{13}\sin\phi_L}{
             s'-m^2_W+im_W\Gamma_W} + \frac{ \cos\phi_L( N_{12}
- 2 Y_q \tan\theta_W N_{11})}{t'-m^2_{\tilde{d}_L}} \nonumber\\
F_R&=&\frac{ 2 N^*_{12} \cos\phi_R- \sqrt{2} N^*_{14}\sin\phi_R}{
             s'-m^2_W+im_W\Gamma_W}
      +\frac{\cos\phi_R( N^*_{12}+ 2 Y_q \tan\theta_W N^*_{11})}{
             u'-m^2_{\tilde{u}_L}}
\end{eqnarray}
where $Y_q=1/6$ is the quark hypercharge. Analogous expressions 
apply to decays into lepton pairs with $Y_l=-1/2$.  
The Mandelstam variables $s'$, $t'$, $u'$ in the form factors
are defined in terms of the 4--momenta of $\chi_1^0, d$ and $\bar{u}$, 
respectively, as
\begin{eqnarray}
s'=(q+\bar{q})^2 \ \ , \ \ t'=(q_0+q)^2 \ \ , \ \ u'=(q_0 +\bar{q})^2 
\end{eqnarray}
while $N$ is the $4\times 4$ matrix rotating  the current neutralino states 
($\tilde{B}, \tilde{W}^3,  \tilde{H}^0_1,  \tilde{H}^0_2$) to the mass
states ($\tilde{\chi}^0_1,.., \tilde{\chi}^0_4$).
The neutralino mass matrix is given by: 
\begin{eqnarray}
{\cal M}_N = \left[ \begin{array}{cccc}
M_1 & 0 & -m_Z s_W \cos\beta & m_Z  s_W \sin\beta \\
0   & M_2 & m_Z c_W \cos\beta & -m_Z  c_W \sin\beta \\
-m_Z s_W \cos\beta & m_Z  c_W \cos\beta & 0 & -\mu \\
m_Z s_W \sin \beta & -m_Z  c_W \sin\beta & -\mu & 0
\end{array} \right]
\end{eqnarray}
Besides the parameters $M_2, \mu$ and $\tan \beta$, which already appear in
the chargino mass matrix, the only additional parameter in
the neutralino mass matrix is $M_1$. [In Grand Unified Theories where the 
gaugino masses are unified at a high-scale, the parameters $M_1$ and $M_2$ 
are related by $M_1= \frac{5}{3} \tan^2 \theta_W M_2$.] \s

The decay distribution of a chargino with  polarization vector $\vec{\cal P}$ 
is formally analogous to the production amplitude after crossing of the
neutralino line and substitution of the generalized charges,  
\begin{eqnarray}
|{\cal D}|^2(n) &=&\frac{4 \pi^2\alpha^2}{s^4_W}\bigg\{
-(t'- m_{\chi_1^\pm}^2)(t- m_{\chi_1^0}^2) |F_L|^2 
-(u'- m_{\chi_1^\pm}^2)(u'- m_{\chi_1^0}^2) |F_R|^2 \nonumber\\
&& -2 m_{\tilde{\chi}^-_1} m_{\tilde{\chi}^0_1} s {\rm Re}(F_L F^*_R)
    \nonumber \\
&& +2 (n\cdot\bar{q})[
 m_{\tilde{\chi}^0_1}(m_{\tilde{\chi}^\pm_1}^2-u'){\rm Re}(F_L F^*_R)
+m_{\tilde{\chi}^\pm_1}(m_{\tilde{\chi}^0  _1}^2-t') |F_L|^2] \nonumber\\
&& -2(n\cdot q)[
 m_{\tilde{\chi}^0_1}(m_{\tilde{\chi}^\pm_1}^2-t'){\rm Re}(F_L F^*_R)
+m_{\tilde{\chi}^\pm_1}(m_{\tilde{\chi}^0  _1}^2-u') |F_R|^2] \bigg\}
\end{eqnarray}
where $n_\mu$ is the $\tilde{\chi}^-_1$ spin 4--vector. 
If the angles in the $f\bar{f}'$ rest system are integrated
out, the $\tilde{\chi}^-_1$ decay final state is described by the energy 
and the polar angle of $\tilde{\chi}^0_1$ [or equivalently 
by the energy and the polar angle of ($f$ {\it plus} $\bar{f}'$)].\s 

For the subsequent discussion of the angular correlations between the two 
charginos in the final state, it is convenient to determine the 
spin--density 
matrix elements $\rho_{\lambda\lambda'}\sim {\cal D}_\lambda {\cal 
D}^*_{\lambda'}$
for the kinematical configuration defined before. Choosing 
the $\tilde{\chi}^\pm_1$ flight direction as quantization axis, the 
spin--density matrix is given by the form 
\begin{eqnarray}
&& \rho_{\lambda\lambda^\prime}
   =\frac{1}{2}\left(\begin{array}{cc}
         1+\kappa\cos\theta^*         &   \kappa\sin\theta^*{\rm e}^{i\phi^*} 
 \\ \kappa\sin\theta^*{\rm e}^{-i\phi^*} &        1-\kappa\cos\theta^*
                     \end{array}\right)\nonumber\\
&& \bar{\rho}_{\bar{\lambda}\bar{\lambda}^\prime}
   =\frac{1}{2}\left(\begin{array}{cc}
 1+\bar{\kappa}\cos\bar{\theta}^* & 
  \bar{\kappa}\sin\bar{\theta}^*{\rm e}^{i\bar{\phi}^*} \\
  \bar{\kappa}\sin\bar{\theta}^*{\rm e}^{-i\bar{\phi}^*} & 
 1-\bar{\kappa}\cos\bar{\theta}^* \end{array}\right)
\end{eqnarray}
$\theta^*$($\bar{\theta}^*$) is the polar angle of the 
$f_1\bar{f}_2$($\bar{f}_3f_4$) system in the 
$\tilde{\chi}^-_1$($\tilde{\chi}^+_1$) rest frame with respect to the original 
flight direction in the laboratory frame, and $\phi^*$($\bar{\phi}^*$) the 
corresponding azimuthal angle with respect to the production plane.
[The orientation of the reference frames has been defined in the 3rd section.] 
The spin analysis--power $\kappa$, which measures the left--right asymmetry,
depends on the final $ud$ or $l\nu$ pair considered in the chargino decays. 
Since left-- and right--chiral form factors $F_L,F_R$,
contribute at the same time, the value of $\kappa$ is determined by the masses 
and couplings of all the particles involved; neglecting effects from
non--zero widths, loops and CP--noninvariant phases, $\kappa$ (and 
$\bar{\kappa}$) is real. While it is important in general to keep the momentum 
dependence of the $W$-propagator, the squark propagators can be approximated
by point propagators; in this case, the analytic expression for $\kappa$
is given by
\begin{eqnarray}
\kappa(s')= 
        - \frac{\beta' (1- \mu_0^2 -2 \mu_h^2) (|F_L|^2-|F_R|^2)}
  {[(1-\mu_0^2)^2
               +\mu_h^2 (1+\mu_0^2-2\mu_h^2)] (|F_L|^2+|F_R|^2)
               -6\mu_0 \mu_h^2 {\rm Re}(F_L F^*_R)} 
\end{eqnarray}
where $\mu_0 = m_{\tilde{\chi}^0_1}/ m_{\tilde{\chi}^-_1}$, $\mu_h = \sqrt{s'} 
/ m_{\tilde{\chi}^-_1}$ and  $\beta = \sqrt{ (1- \mu_0^2 -\mu_h^2)^2 - 
4\mu_h^2 \mu_0^2}$. 
Characteristic examples for $\kappa(s')$, without using the point-propagator
approximations, are presented in Fig.~6 for the same choice of parameters as 
Fig.~2b; the squark masses are set to 300~GeV, and the 
gaugino masses are assumed universal at the unification scale. 
The size of $\kappa$ decreases as the invariant mass of the
fermion system increases. Good reconstruction of the two--fermion
system with a modest invariant mass is therefore required to make efficient 
use of the polarization observables [and to make a precise determination of 
the end point of the invariant mass spectrum, which gives the neutralino 
mass].\s

\subsubsection*{4.2 Angular Correlations}
\label{sec:angular correlation}

Since the $\tilde{\chi}^\pm_1$ lifetime is very small, only the correlated
production and decay can be observed experimentally:
\begin{center}
\begin{picture}(300,100)(0,0)
\Text(130,80)[]{$e^+e^-\rightarrow\tilde{\chi}^+_1\tilde{\chi}^-_1$}
\Line(145,70)(145,40)
\Line(155,70)(155,60)
\Text(155,61)[l]{$\rightarrow\tilde{\chi}^0_1+(f_1\bar{f}_2)$}
\Text(145,41)[l]{$\longrightarrow\hskip 0.15cm\tilde{\chi}^0_1+(\bar{f}_3f_4)$}
\end{picture}
\end{center}
\vskip -1cm
The analysis is complicated as the two invisible neutralinos in the final state
do not allow for a complete reconstruction of the events. In particular,
it is not possible to measure the $\tilde{\chi}^\pm_1$ production angle
$\Theta$; this angle can be determined only up to a two--fold ambiguity. \s

In covariant language the final state distributions are found by combining 
the polarized cross section
\begin{eqnarray}
{\rm d}\sigma=\langle{\rm d}\sigma\rangle\frac{1}{4}
       \left[1-{\cal P}^\mu n_\mu-\bar{\cal P}^\mu \bar{n}_\mu
              +{\cal Q}^{\mu\nu} n_\mu\bar{n}_\nu\right]
\end{eqnarray}
with the polarized decay distributions
\begin{eqnarray}
&&{\rm d}\Gamma=\langle{\rm d}\Gamma\rangle
                \left[1-{\cal P}^{'\mu}n_\mu\right] \nonumber\\
&&{\rm d}\bar{\Gamma}=\langle{\rm d}\bar{\Gamma}\rangle
                \left[1-\bar{\cal P}^{'\mu}\bar{n}_\mu\right] 
\end{eqnarray}
Inserting the completeness relations
\begin{eqnarray}
\sum n_\mu n_\nu=-g_{\mu\nu}+ k_\mu k_\nu / m^2_{\tilde{\chi}^-_1}
= \eta_{\mu \nu} \ \ 
\hskip 0.5cm  [{\rm etc}]
\end{eqnarray}
the overall event topology can be calculated from the formula
\begin{eqnarray}
{\rm d}\sigma_{\rm final}&=&\langle{\rm d}\sigma\rangle 
   \langle{\rm d}\Gamma\rangle\langle{\rm d}\bar{\Gamma}\rangle\frac{1}{4}
   \bigg[1+\eta_{\mu\alpha}{\cal P}^\mu{\cal P}^{'\alpha}
      +\bar{\eta}_{\nu\beta}\bar{\cal P}^\nu\bar{\cal P}^{'\beta}
      +\eta_{\mu\alpha}\bar{\eta}_{\nu\beta}
          {\cal Q}^{\mu\nu}{\cal P}^{'\alpha}\bar{\cal P}^{'\beta}\bigg]
\end{eqnarray}
with covariant expressions for ${\cal P}_\mu$ {\it etc} as noticed
earlier. This formula provides the basis for deriving
any distribution or correlation between the final state particles. \s

Alternatively we may choose the helicity analysis to interpret
the event topology. Denoting the matrix elements 
${\cal M}=\sum T_{\lambda\bar{\lambda}}{\cal D}_\lambda
                         \overline{\cal D}_{\bar{\lambda}}$, 
the 7--fold differential cross section can be derived from the transition 
probability $|{\cal M}|^2=\sum T_{\lambda\bar{\lambda}}
T^*_{\lambda'\bar{\lambda}'}\rho_{\lambda\lambda'}
\bar{\rho}_{\bar{\lambda}\bar{\lambda}'}$: 
\begin{eqnarray}
&& \frac{{\rm d}^7\sigma(e^+e^-\rightarrow\tilde{\chi}^-_1\tilde{\chi}^+_1
               \rightarrow \tilde{\chi}^0_1\tilde{\chi}^0_1(f_1\bar{f}_2)
               (\bar{f}_3f_4))}{{\rm d}\cos\Theta {\rm d}s'
               {\rm d}\cos\theta^* {\rm d}\phi^*{\rm d}\bar{s}'
               {\rm d}\cos\bar{\theta}^* {\rm d}\bar{\phi}^*} \nonumber \\
&& \hskip 1.5cm  = \frac{\alpha^2\beta}{124\pi s} 
   {\rm Br}(\tilde{\chi}^-\rightarrow\tilde{\chi}^0_1f_1\bar{f}_2)
   {\rm Br}(\tilde{\chi}^+\rightarrow\tilde{\chi}^0_1\bar{f}_3f_4)
   \Sigma(\Theta;s',\theta^*,\phi^*;
          \bar{s}',\bar{\theta}^*,\bar{\phi}^*) \phantom{HALLO} 
\end{eqnarray}
with
\begin{eqnarray}
\Sigma=\sum_{\lambda\bar{\lambda}}\sum_{\lambda'\bar{\lambda}'}\sum_\sigma
       \langle\sigma;\lambda\bar{\lambda}\rangle
       \langle\sigma;\lambda'\bar{\lambda}'\rangle^*
       \rho_{\lambda\lambda'}\bar{\rho}_{\bar{\lambda}\bar{\lambda}'}
\end{eqnarray}
The unobservable $\tilde{\chi}^\pm_1$ production angle $\Theta$ will be 
integrated out and, for the sake of simplicity, the $(f_1\bar{f}_2)$ and 
$(\bar{f}_3f_4)$ invariant masses $\sqrt{s'}$, $\sqrt{\bar{s}'}$ too.
The integrated cross section
\begin{eqnarray}
&&\frac{{\rm d}^4\sigma(e^+e^-\rightarrow\tilde{\chi}^-_1\tilde{\chi}^+_1
               \rightarrow \tilde{\chi}^0_1\tilde{\chi}^0_1(f_1\bar{f}_2)
               (\bar{f}_3f_4))}{{\rm d}\cos\theta^*{\rm d}\phi^*
               {\rm d}\cos\bar{\theta}^*{\rm d}\bar{\phi}^*}\nonumber\\
&& \hskip 1.5cm =\frac{\alpha^2\beta}{124\pi s} 
   {\rm Br}(\tilde{\chi}^-\rightarrow\tilde{\chi}^0_1f_1\bar{f}_2)
   {\rm Br}(\tilde{\chi}^+\rightarrow\tilde{\chi}^0_1\bar{f}_3f_4)
    \Sigma(\theta^*,\phi^*;\bar{\theta}^*,\bar{\phi}^*)
\phantom{HALLOOOO}
\end{eqnarray}
can be decomposed into sixteen independent angular parts
\begin{eqnarray}
\Sigma&=&\Sigma_{\rm unpol}+\cos\theta^*\kappa{\cal P}
       +\cos\bar{\theta}^* \bar{\kappa}\bar{\cal P}
       +\cos\theta^*\cos\bar{\theta}^*\kappa\bar{\kappa}{\cal Q}\nonumber\\
     &&+\sin\theta^*\cos\phi^*\kappa{\cal U}
       +\sin\theta^*\sin\phi^*\kappa\bar{\cal U}\nonumber\\
     &&+\sin\bar{\theta}^*\cos\bar{\phi}^*\bar{\kappa}{\cal V}
       +\sin\bar{\theta}^*\sin\bar{\phi}^*\bar{\kappa}\bar{\cal V}\nonumber\\
     &&+\sin\theta^*\cos\phi^*\cos\bar{\theta}^*\kappa\bar{\kappa}{\cal W}
       +\sin\theta^*\sin\phi^*\cos\bar{\theta}^*\kappa\bar{\kappa}\bar{\cal W}
        \nonumber\\
     &&+\cos\theta^*\sin\bar{\theta}^*\cos\bar{\phi}^*\kappa\bar{\kappa}
        {\cal X}
       +\cos\theta^*\sin\bar{\theta}^*\sin\bar{\phi}^*\kappa\bar{\kappa}
        \bar{\cal X}\nonumber\\
     &&+\sin\theta^*\sin\bar{\theta}^*\cos(\phi^*+\bar{\phi}^*)
        \kappa\bar{\kappa}{\cal Y}
       +\sin\theta^*\sin\bar{\theta}^*\sin(\phi^*+\bar{\phi}^*)
        \kappa\bar{\kappa}\bar{\cal Y}\nonumber\\
     &&+\sin\theta^*\sin\bar{\theta}^*\cos(\phi^*-\bar{\phi}^*)
        \kappa\bar{\kappa}{\cal Z}
       +\sin\theta^*\sin\bar{\theta}^*\sin(\phi^*-\bar{\phi}^*)
        \kappa\bar{\kappa}\bar{\cal Z} 
\end{eqnarray}
The sixteen coefficients are combinations of helicity amplitudes, 
corresponding 
to the unpolarized cross section, $2 \times 3$ polarization components and 
$3 \times 3$ spin--spin correlations. 

\bigskip 

\nn (i) \underline{Unpolarized cross section:} 
\begin{eqnarray}
\Sigma_{\rm unpol}&=&\frac{1}{4}\int {\rm d}\cos\Theta\sum_{\sigma=\pm}
      \bigg[|\langle\sigma;++\rangle|^2+|\langle\sigma;+-\rangle|^2
           +|\langle\sigma;-+\rangle|^2+|\langle\sigma;--\rangle|^2
      \bigg]
\end{eqnarray}

\smallskip

\nn (ii) \underline{Polarization components:} 
\begin{eqnarray}
{\cal P}&=&\frac{1}{4}\int{\rm d}\cos\Theta\sum_{\sigma=\pm}
      \bigg[|\langle\sigma;++\rangle|^2+|\langle\sigma;+-\rangle|^2
           -|\langle\sigma;-+\rangle|^2-|\langle\sigma;--\rangle|^2
      \bigg] \nonumber\\
\bar{\cal P}&=&\frac{1}{4}\int {\rm d}\cos\Theta\sum_{\sigma=\pm}
      \bigg[|\langle\sigma;++\rangle|^2+|\langle\sigma;-+\rangle|^2
           -|\langle\sigma;+-\rangle|^2-|\langle\sigma;--\rangle|^2
      \bigg] \nonumber\\
{\cal U}&=&\frac{1}{2}\int {\rm d}\cos\Theta\sum_{\sigma=\pm}
      {\rm Re}\bigg\{\langle\sigma;-+\rangle\langle\sigma;++\rangle^*
                    +\langle\sigma;--\rangle\langle\sigma;+-\rangle^*\bigg\}
       \nonumber\\
{\cal V}&=&\frac{1}{2}\int {\rm d}\cos\Theta\sum_{\sigma=\pm}
      {\rm Re}\bigg\{\langle\sigma;+-\rangle\langle\sigma;++\rangle^*
                    +\langle\sigma;--\rangle\langle\sigma;-+\rangle^*\bigg\}
\end{eqnarray}
and $\bar{\cal U}, \bar{\cal V}$ defined as ${\cal U}, {\cal V}$
after replacing Re by Im. 

\bigskip 

\nn (iii) \underline{Spin--spin correlations:}  
\begin{eqnarray}
{\cal Q}&=&\frac{1}{4}\int {\rm d}\cos\Theta\sum_{\sigma=\pm}
      \bigg[|\langle\sigma;++\rangle|^2-|\langle\sigma;+-\rangle|^2
           -|\langle\sigma;-+\rangle|^2+|\langle\sigma;--\rangle|^2
      \bigg]\nonumber\\
{\cal W}&=&\frac{1}{2}\int {\rm d}\cos\Theta\sum_{\sigma=\pm}
      {\rm Re}\bigg\{\langle\sigma;-+\rangle\langle\sigma;++\rangle^*
                    -\langle\sigma;--\rangle\langle\sigma;+-\rangle^*\bigg\}
       \nonumber\\
{\cal X}&=&\frac{1}{2}\int {\rm d}\cos\Theta\sum_{\sigma=\pm}
      {\rm Re}\bigg\{\langle\sigma;+-\rangle\langle\sigma;++\rangle^*
                    -\langle\sigma;--\rangle\langle\sigma;-+\rangle^*\bigg\}
       \nonumber\\
{\cal Y}&=&\frac{1}{2}\int {\rm d}\cos\Theta\sum_{\sigma=\pm} 
      {\rm Re}\bigg\{\langle\sigma;--\rangle\langle\sigma;++\rangle^*\bigg\}
       \nonumber\\
{\cal Z}&=&\frac{1}{2}\int {\rm d}\cos\Theta\sum_{\sigma=\pm} 
      {\rm Re}\bigg\{\langle\sigma;-+\rangle\langle\sigma;+-\rangle^*\bigg\}
\end{eqnarray}
and $\bar{\cal W}, \bar{\cal X}, \bar{\cal Y}, \bar{\cal Z}$ defined  
as ${\cal W}, {\cal X}, {\cal Y}, {\cal Z}$ after replacing Re by Im. \s

Since loops and the width of the $Z$--boson can be neglected for high 
energies, 
the helicity amplitudes in eq.~(\ref{eq:helicity amplitude}) can 
be taken real in CP--invariant theories. 
In this approximation the six functions 
$\bar{\cal U}, \bar{\cal V}, \bar{\cal W}, \bar{\cal X},
\bar{\cal Y}, \bar{\cal Z}$ can be discarded. Moreover, from 
CP--invariance, $ \langle\sigma;\lambda\bar{\lambda}\rangle
  =-(-)^{(\lambda-\bar{\lambda})}\langle\sigma;-\bar{\lambda}
  -\lambda\rangle $,  
it follows that $\bar{\cal P}=-{\cal P}$, ${\cal U}=-{\cal V}$ and ${\cal 
W}={\cal X}$. The overall topology is therefore determined by seven 
independent functions: 
$\Sigma_{\rm unpol}, {\cal P}, {\cal Q}, {\cal U},
{\cal W}, {\cal Y}, {\cal Z}$. \s

In terms of the generalized charges, the correlation functions 
${\cal Q}$ and ${\cal Y}$, which we will discuss next in 
detail,  are given by
\begin{eqnarray}
&& {\cal Q}=-4\int {\rm d}\cos\Theta\left[(\beta^2+\cos^2\Theta)Q_1
         +(1-\beta^2)\cos^2\Theta Q_2+2\beta\cos\Theta Q_3\right]\nonumber \\
&& {\cal Y}=-2\int {\rm d}\cos\Theta(1-\beta^2)\left[Q_1+Q_2\right]\sin^2
  \Theta 
\end{eqnarray}

The observables ${\cal P}$, $\bar{\cal P}$, ${\cal Q}$ and ${\cal Y}$
enter into the cross section together with the spin analysis-power 
$\kappa (\bar{\kappa})$. CP--invariance leads to the relation
$\bar{\kappa}=-\kappa$. Therefore, taking the ratios ${\cal P}^2/{\cal Q}$
and ${\cal P}^2/{\cal Y}$, these unknown quantities can be eliminated
so that the two ratios reflect unambiguously the properties of the chargino
system, being not affected by the neutralinos. It is thus possible to study 
the chargino sector in isolation by measuring the mass of the lightest
chargino, the total production cross section and the spin(--spin) correlations.
The energy dependence of the two ratios ${\cal P}^2/{\cal Q}$ and
${\cal P}^2/{\cal Y}$ is shown in Fig.~7; the same parameters are chosen
as in the previous figures. The two ratios are sensitive to the quartic charges
at sufficiently large c.m. energies since the charginos are, on the average,
unpolarized at the threshold, c.f. eqs.~(24).
Note that ${\cal Y}$ vanishes for asymptotic energies so that an optimal
energy must be chosen not far above threshold to measure this observable.

\subsection*{5. Observables and Extraction of SUSY Parameters}
\label{sec:observable}

The pair production of the lightest chargino $\chi_1^\pm$ is 
characterized by the chargino mass $m_{\tilde{\chi}^\pm_1}$, the 
sneutrino mass $m_{\tilde{\nu}}$, and the two mixing 
angles $\cos 2\phi_{L,R}$. These three quantities
can be determined from the production cross section and the 
spin correlations. \s

The mass $m_{\tilde{\chi}^\pm_1}$ can be measured very precisely near 
the threshold where the production cross section 
$\sigma(e^+e^-\rightarrow \tilde{\chi}^+_1\tilde{\chi}^-_1)$ rises 
sharply with the velocity $\beta=\sqrt{1-4m_{\chi_1^\pm}^2/s}$.
Combining the energy variation of the cross section
with the measurement of the spin correlations, the sneutrino mass 
$m_{\tilde{\nu}}$ and the two mixing angles $\cos 2\phi_L$ and 
$\cos 2\phi_R$ can be extracted. \s

The decay angles $\{\theta^*,\phi^*\}$ and $\{\bar{\theta}^*,\bar{\phi}^*\}$,
which are used to measure the $\chi_1^\pm$ chiralities, 
are defined in the rest frame of the charginos $\tilde{\chi}^-_1$
and $\tilde{\chi}^+_1$, respectively. Since there are two invisible
neutralinos in the final state, they can not be reconstructed completely.
However, the longitudinal components and the inner product of the transverse
components can be reconstructed from the momenta measured in the laboratory
frame (see e.g. Ref.~\cite{R11B}), 
\begin{eqnarray}
&&\cos\theta^*=\frac{1}{\beta |\vec{p}^*|} \left(\frac{E}{\gamma}-E^*\right)
\ \ , \ \  \cos\bar{\theta}^*=\frac{1}{\beta |\vec{\bar{p}}^*|}
               \left(\frac{\bar{E}}{\gamma}-\bar{E}^*\right) \nonumber\\
&&\sin\theta^*\sin\bar{\theta}^*\cos(\phi^*+\bar{\phi}^*)=
         \frac{|\vec{p}||\vec{\bar{p}}|}{|\vec{p}^*||\vec{\bar{p}}^*|}
         \cos\vartheta+\frac{\left(E-E^*/\gamma \right)
           \left(\bar{E}-\bar{E}^*/\gamma\right)}{\beta^2
           |\vec{p}^*||\vec{\bar{p}}^*|}
\end{eqnarray}
where $\gamma=\sqrt{s}/{2m_{\tilde{\chi}^\pm_1}}$. $E (\bar{E})$ and 
$E^*(\bar{E}^*)$ are the energies of the two hadronic systems in the laboratory
frame and in the rest frame of the charginos, respectively; 
$\vec{p} (\vec{\bar{p}})$ and $\vec{p}^* (\vec{\bar{p}^*})$ are 
the corresponding
momenta. $\vartheta$ is the angle between the momenta of the two hadronic 
systems; 
the angle between the vectors in the transverse plane is given by 
$\Delta \phi^* = 2\pi-(\phi^*+\bar{\phi}^*)$ for the reference frames 
defined earlier. The polarization 
and correlation
functions, ${\cal P}$, ${\cal Q}$ and ${\cal Y}$ can  therefore be
measured directly. 
Since the polarization ${\cal P}$ is odd under parity and charge--conjugation, 
it is necessary to identify the chargino electric charges in this case. 
This can be accomplished by making use of the mixed 
leptonic and hadronic decays of the chargino pairs. On the other hand, the
observables ${\cal Q}$ and 
${\cal Y}$ are defined without charge identification so that the 
dominant hadronic decay modes of the charginos can be exploited. \s

The measurements of the cross section at an energy $\sqrt{s}$, 
and either of the ratios ${\cal P}^2/{\cal Q}$ or ${\cal P}^2/{\cal Y}$ 
can be interpreted as 
contour lines in the plane $\{\cos 2\phi_L,\cos 2\phi_R\}$ 
which intersect with large angles so that a high precision in the 
resolution can be achieved. A representative example for the 
determination of $\cos 2\phi_L$ and $\cos 2\phi_R$ is shown in 
Fig.~8. The mass of the light chargino is set to $m_{\tilde{\chi}^\pm_1}=95$ 
GeV, and the ``measured'' cross section, ${\cal P}^2/{\cal Q}$ and 
${\cal P}^2/{\cal Y}$  are taken to be 
\begin{eqnarray}
\sigma(e^+e^-\rightarrow\tilde{\chi}^+_1\tilde{\chi}^-_1)=0.37\ \ 
{\rm pb},\qquad
\frac{{\cal P}^2}{\cal Q}=-0.24,\qquad 
\frac{{\cal P}^2}{\cal Y}= -3.66
\label{eq:measured}
\end{eqnarray}
at $\sqrt{s}=500$ GeV.
The three contour lines meet at a single point $\{\cos 2\phi_L,\cos 2\phi_R\}
=\{-0.8,-0.5\}$ for $m_{\tilde{\nu}}=250$ GeV; note that
the sneutrino mass can be determined together with the mixing angles 
from the ``measured values" in eq.~(\ref{eq:measured}).\s

The solutions can be discussed most transparently by introducing the two 
triangular quantities
\begin{eqnarray}
p=\cot(\phi_R-\phi_L)\ \ {\rm and}\  \ q=\cot(\phi_R+\phi_L)
\end{eqnarray}
They can be derived from the measured values $\cos 2\phi_L$ and
$\cos 2\phi_R$ up to a discrete ambiguity which is due to the sign 
ambiguity in $\sin 2\phi_L$ and $\sin 2\phi_R$.
Solving the set
\begin{eqnarray}
p^2+q^2 &=&\frac{2(\sin^2 2\phi_L+\sin^2 2\phi_R)}{(\cos 2\phi_L-\cos 2
\phi_R)^2}
           \nonumber\\
  pq    &=&\frac{\cos 2\phi_L+\cos 2\phi_R}{\cos 2\phi_L-\cos 2\phi_R}
\nonumber\\
p^2-q^2 &=&\frac{4\sin 2\phi_L\sin 2\phi_R}{(\cos 2\phi_L-\cos 2\phi_R)^2} 
\end{eqnarray}
the solutions $(p,q)$ in point (1) and point (2) of Fig.~8 are found 
for $\sin 2\phi_L \sin 2\phi_R   
\stackrel{\textstyle >}{<} 0$, respectively.
A second set is found by reversing the signs of the solutions pairwise.
These solutions are shown for positive values of $pq$ in Fig.~8. \s

{}From the solutions $(p,q)$ derived above, the SUSY parameters can be 
determined in the following way.\s

\nn {\bf (i) \underline{$\tan\beta$:}} Depending on the relative magnitude of 
$\cos 2\phi_R$ and $\cos 2\phi_L$, the value of $\tan\beta$ is either
larger or smaller than unity. The first case is realized for
\begin{eqnarray}
\cos 2\phi_R > \cos 2\phi_L\ \ : \ \ 
\tan\beta = \frac{p^2-q^2\pm 2\sqrt{\chi^2(p^2+q^2+2-\chi^2)}}{
       (\sqrt{1+p^2}-\sqrt{1+q^2})^2-2\chi^2} 
      \; \Rightarrow \; \tan\beta\geq 1
\end{eqnarray}
where $\chi^2 =m^2_{\tilde{\chi}^\pm_1}/m^2_W$.
If the denominator is positive, there are either up to two solutions
for $\tan\beta > 1$ in point (1) and none in point (2), or
at most one in point (1) and at most one in point (2). The possible
solutions can be counted in an analogous way if the denominator is
negative; the r\^{o}les of point (1) and point (2) are just 
interchanged. The same counting is also valid in the second case
for
\begin{eqnarray}
\cos 2\phi_R < \cos 2\phi_L\ \ : \ \ 
\tan\beta = \frac{(\sqrt{1+p^2}-\sqrt{1+q^2})^2-2\chi^2}{ 
             p^2-q^2\pm 2\sqrt{\chi^2(p^2+q^2+2-\chi^2)}}
      \; \Rightarrow \; \tan\beta\leq 1
\end{eqnarray}
Thus, only a two--fold ambiguity is inferred from all the solutions in 
point (1) and point (2).\s

\nn {\bf (ii) \underline{$M_2,\mu$:}} 
The gaugino and higgsino mass parameters are given in terms of $p$ and 
$q$ by the relations
\begin{eqnarray}
M_2&=&\frac{m_W}{\sqrt{2}}\bigg[(p+q)\sin\beta-(p-q)\cos\beta\bigg]
       \nonumber\\
\mu&=&\frac{m_W}{\sqrt{2}}\bigg[(p-q)\sin\beta-(p+q)\cos\beta\bigg]
\label{eq:M2_mu}
\end{eqnarray}
The parameters $M_2$, $\mu$ are uniquely fixed if $\tan\beta$ is chosen 
properly in point (1) and/or point (2). Since $\tan\beta$ is invariant
under pairwise reflection of the signs in $(p,q)$, the definition 
$M_2 > 0$ can be exploited to remove this additional ambiguity.\s

As a result, the fundamental SUSY parameters [$\tan\beta;M_2,\mu$] 
can be derived from the observables $m_{\tilde{\chi}^\pm_1}$
and $\cos 2\phi_R$, $\cos 2\phi_L$ up to at most a two--fold ambiguity.\s

Returning to the ``experimental values" of mass, cross section and spin
correlations introduced above, the following SUSY parameters are  
extracted:
\begin{eqnarray}
{\rm Point\ \ (2)} \;  : \; [\tan\beta; M_2,\mu] =
        \left\{\begin{array}{l}
              [1.06; \;  83{\rm GeV}, \; -59{\rm GeV}] \\
                  { }\\ {}
              [3.33; \;  248{\rm GeV}, \; 123{\rm GeV}]
              \end{array}\right.
\end{eqnarray}
Two solutions are derived from the ``experimental values" in this case;
point (1) gives negative values for $\tan\beta$.
In practice, the errors in the observables $m_{\tilde{\chi}^\pm_1}$
and $\cos 2\phi_{R,L}$ must be analyzed experimentally and the migration
to the fundamental SUSY parameters must be studied properly. This however
is beyond the scope of the purely theoretical analysis presented in this 
paper.

\subsection*{6. Conclusions}
\label{sec:conclusion}

We have analyzed how the parameters of the chargino system, 
the mass of the lightest chargino $m_{\tilde{\chi}^\pm_1}$
and the size of the wino and higgsino components in the chargino 
wave--functions,  
parameterized in terms of the two angles $\phi_L$ and $\phi_R$, can
be extracted from pair production of the lightest chargino state
in $\ee$ annihilation. 
In addition to the total production cross section, angular correlations
among the chargino decay products give rise to two independent observables 
which can be measured directly despite of the two invisible neutralinos
in the final state. \s

{}From the chargino mass $m_{\tilde{\chi}^\pm_1}$ and the two mixing angles
$\phi_L$ and $\phi_R$, the fundamental supersymmetric parameters $\tan\beta$, 
$M_2$ and $\mu$ can be extracted up to at most a two-fold discrete ambiguity. 
Moreover, from the energy distribution of the final particles
in the decay of the chargino, the mass of the lightest neutralino can be 
measured; this allows to determine the parameter $M_1$ so that 
also the neutralino mass matrix can be reconstructed
in a model-independent way. \s

The analysis has been carried out for scenarios in which the chargino 
sector is  CP--invariant. 
The generalization to CP non--invariant theories \cite{R11, R12} 
is touched upon
in a brief appendix for completeness. \bigskip \bigskip

\subsubsection*{Acknowledgments}

This work was supported by the KOSEF-DFG large collaboration
project, Project No. 96-0702-01-01-2, and by the Polish State Committee
for Scientific Research, Grant No. 2 P 03B 030 14. Thanks go to
M.~Raidal for comments on the manuscript.

\bigskip \bigskip

\subsection*{APPENDIX: Complex mass parameters}

In {\cal CP}--noninvariant theories, the gaugino mass $M_2$ and the Higgs mass 
parameter $\mu$ can be complex. However, by 
reparametrizations of the fields, $M_2$ can be assumed real and positive
without loss of generality \cite{R11} so that the only 
non--trivial invariant phase is attributed to $\mu$:
\begin{equation}
\mu = |\mu| {\rm e}^{i\theta}
\end{equation}
In these theories the complex chargino mass matrix (\ref{eq:mass matrix})
is diagonalized by two unitary matrices $U_L$ and $U_R$:
\begin{eqnarray}
U_{L,R}
 \left(\begin{array}{c}
      \tilde{W}^- \\
      \tilde{H}^-
       \end{array}\right)_{L,R}=
 \left(\begin{array}{c}
      \tilde{\chi}^-_1 \\
      \tilde{\chi}^-_2
       \end{array}\right)_{L,R}
\end{eqnarray}
They can be parameterized in the following way: 
\begin{eqnarray}
&& U_L=
      \left(\begin{array}{cc}
     \cos\phi_L                     &  {\rm e}^{-i\beta_L}\sin\phi_L \\
      -{\rm e}^{i\beta_L}\sin\phi_L  &  \cos\phi_L
      \end{array}\right) \nonumber \\    
&& U_R=\left(\begin{array}{cc}
     {\rm e}^{i\gamma_1}   &    0   \\
                   0       &  {\rm e}^{i\gamma_2}
             \end{array}\right)
      \left(\begin{array}{cc}
     \cos\phi_R                     &  {\rm e}^{-i\beta_R}\sin\phi_R \\
      -{\rm e}^{i\beta_R}\sin\phi_R  &  \cos\phi_R
      \end{array}\right)  
\label{eq:mixing matrix}
\end{eqnarray}
The eigenvalues $m^2_{\tilde{\chi}^\pm_{1}}$ involve the angle $\theta$:
\begin{eqnarray}
m^2_{\tilde{\chi}^\pm_{1,2}}=\frac{1}{2}\bigg[M_2^2+|\mu|^2+2m^2_W
      \mp\Delta_C\bigg]
\end{eqnarray}
with 
\begin{eqnarray}
\Delta_C= \sqrt{(M_2^2+|\mu|^2+2m^2_W)^2-4(M_2^2|\mu|^2
         -2m^2_W M_2|\mu|\sin 2\beta\cos\theta+m^4_W\sin^2 2\beta)}
\end{eqnarray}
The four nontrivial phase angles $\{\beta_L,\beta_R,\gamma_1,\gamma_2\}$ 
also depend on the invariant angle $\theta$:
\begin{eqnarray}
\tan\beta_{L}=-\frac{\sin\theta}{\cos\theta
               +\frac{M_2}{|\mu|}\cot\beta} 
& , & \tan\beta_R=+\frac{\sin\theta}{\cos\theta
               +\frac{M_2}{|\mu|}\tan\beta}\nonumber\\
\tan\gamma_1=+\frac{\sin\theta}{\cos\theta+\frac{M_2}{|\mu|}
           \frac{m^2_{\tilde{\chi}^\pm_1}-|\mu|^2}{m^2_W\sin 2\beta}}
&,& \tan\gamma_2=-\frac{\sin\theta}{\cos\theta+\frac{M_2}{|\mu|}
           \frac{m^2_W\sin 2\beta}{m^2_{\tilde{\chi}^\pm_2}-M_2^2}}
\end{eqnarray}
The mixing angles $\phi_{LR}$ are given by the relations

\begin{eqnarray}
&&\cos 2\phi_L=-\frac{M_2^2-|\mu|^2-2m^2_W\cos 2\beta}{\Delta_C}
              \nonumber\\ 
&&\sin 2\phi_L=-\frac{2m_W\sqrt{M_2^2+|\mu|^2+(M_2^2-|\mu|^2)\cos 2\beta
               + 2M_2|\mu|\sin 2\beta\cos\theta}}{\Delta_C}\nonumber
\end{eqnarray}
and
\begin{eqnarray}
&&\cos 2\phi_R=-\frac{M_2^2-|\mu|^2+2m^2_W\cos 2\beta}{\Delta_C}
              \nonumber\\ 
&&\sin 2\phi_R=-\frac{2m_W\sqrt{M_2^2+|\mu|^2-(M_2^2-|\mu|^2)\cos 2\beta
               + 2M_2|\mu|\sin 2\beta\cos\theta}}{\Delta_C}
\end{eqnarray}

The $\tilde{\chi}_1 \tilde{\chi}_1 \gamma$ and 
$\tilde{\chi}_1 \tilde{\chi}_1 Z$ vertices are real and they can 
be expressed by the mixing angles $\phi_{L,R}$ in the same way 
as in {\cal CP}--invariant
theories. Even though the new phases enter the $e\tilde{\nu} \tilde{\chi}_1$
vertex, they do not affect the $\tilde{\nu}$--exchange amplitude. 
As a result, the 
analytical expressions of all observables in the diagonal process 
$e^+e^- \rightarrow \tilde{\chi}^+_1 \tilde{\chi}^-_1$ 
remain the same when described in terms of the mixing angles $\phi_L$ 
and $\phi_R$. Since the density matrix (31) is factored out completely and the
form of 
the sixteen coefficients in eqs.~(41), (42) and (43) does not change
by CP-noninvariance, the analysis described in this paper 
is not changed. \s

If the phase $\theta$ is introduced, the observables 
$m_{\tilde{\chi}^{\pm}_1}, \phi_L$ and $\phi_R$ are insufficient to
reconstruct the fundamental SUSY parameters $\tan \beta$, $M_2$, $|\mu|$
and $\theta$ {\it in toto}. In this complex situation, one more observable
is needed. The additional information may be extracted, for example,
from the $\tilde{\chi}^\pm_2$ mass. [Else the neutralino system may
be exploited to provide the additional observable
\cite{R12}]. The {\cal CP}--odd phase $\theta$ can be determined
directly in the non--diagonal process
$e^+e^- \rightarrow \tilde{\chi}^+_1 \tilde{\chi}^-_2$, see Ref.~\cite{R11}. 

\bigskip \bigskip

\newpage
\mbox{ }
\vskip  1cm
\addtocounter{figure}{1}

\begin{center}
\begin{figure}[htb]
\hbox to\textwidth{\hss\epsfig{file=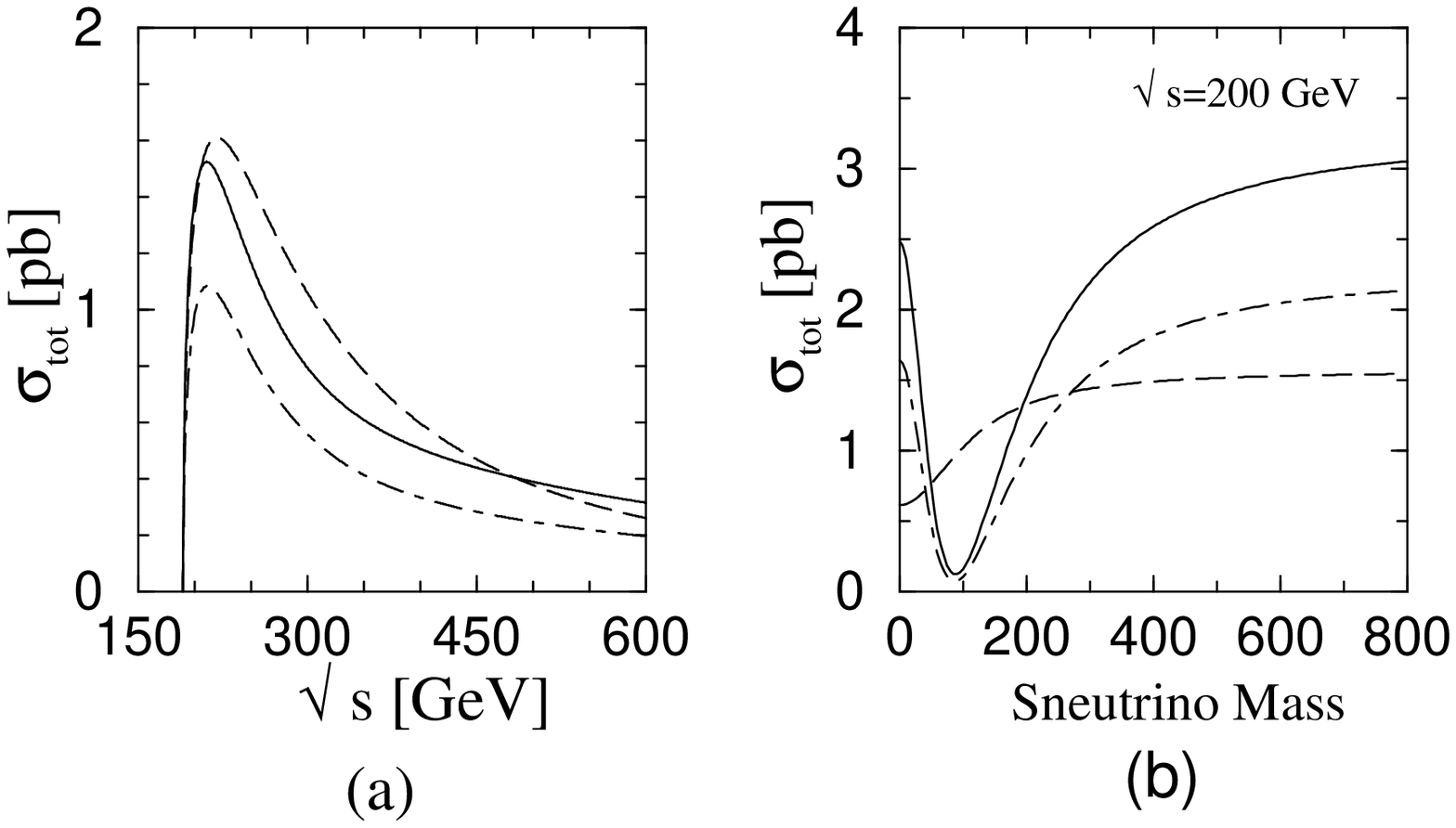,height=6cm}\hss}
\caption{\it The  cross section for the production of 
             charginos, (a) as a function of 
             the c.m. energy with $m_{\tilde{\nu}}=200$ GeV, and (b) as a 
             function of the sneutrino mass with 
             $\protect\sqrt{s}=200$ GeV for the 
             representative set of SUSY parameters 
             in eq.~(\ref{eq:parameter}): 
             solid line  for the gaugino case, dashed line for the
             higgsino case, and  dot-dashed line for the mixed case.}
\label{fig:snu}
\end{figure}
\end{center}

\vskip -1cm


%
\begin{center}
\begin{figure}[htb]
\hbox to\textwidth{\hss\epsfig{file=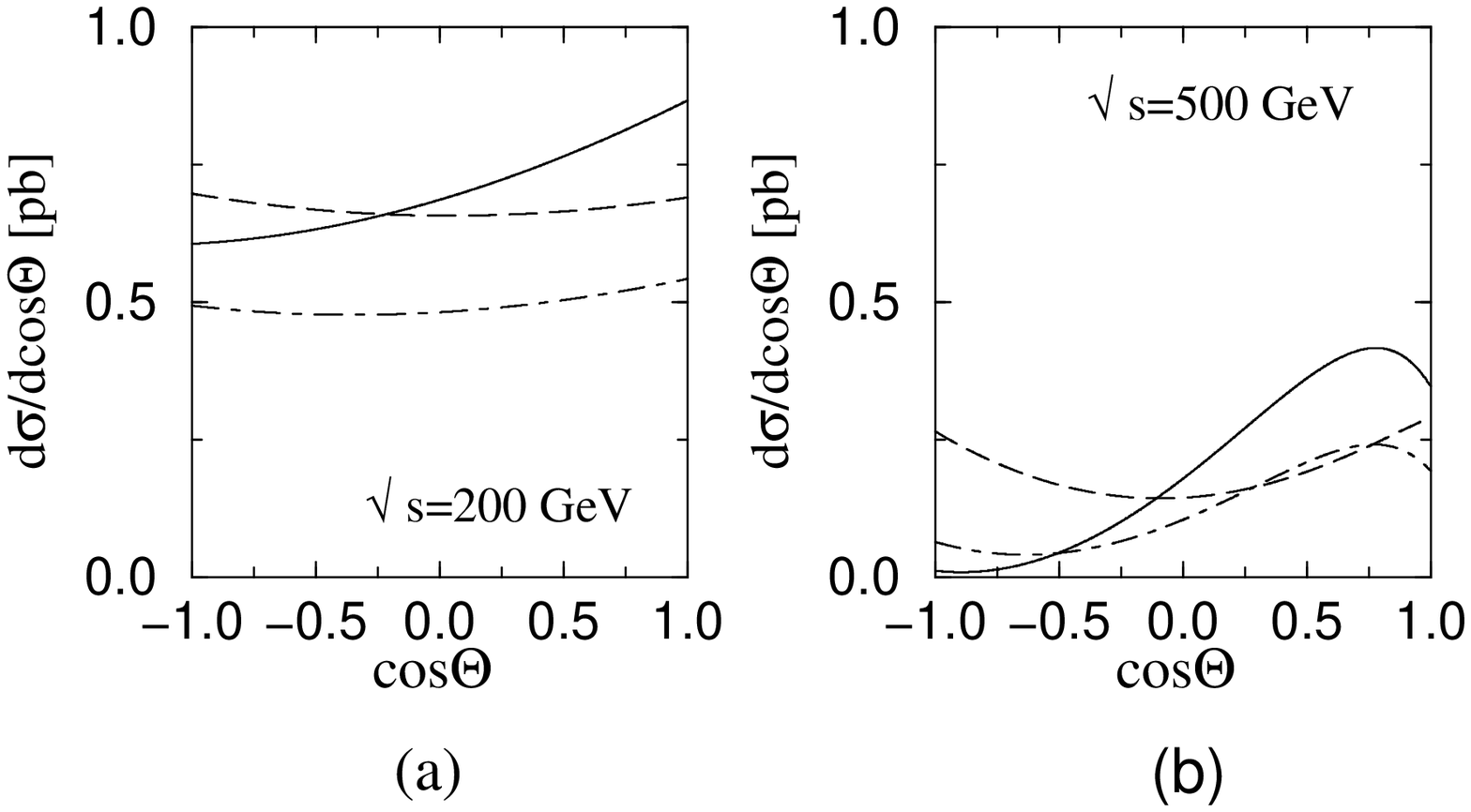,height=6cm}\hss}
\caption{\it The angular distribution as a function of the scattering angle
             at  c.m. energies  (a) 200 and (b) 500 GeV  
              for the set of SUSY parameters 
             in eq.~(\ref{eq:parameter}) and $m_{\tilde{\nu}}=200$ GeV.}
\label{fig:xrs}
\end{figure}
\end{center}

\newpage
\mbox{ }
\vskip 3cm

\begin{center}
\begin{figure}[htb]
\hbox to\textwidth{\hss\epsfig{file=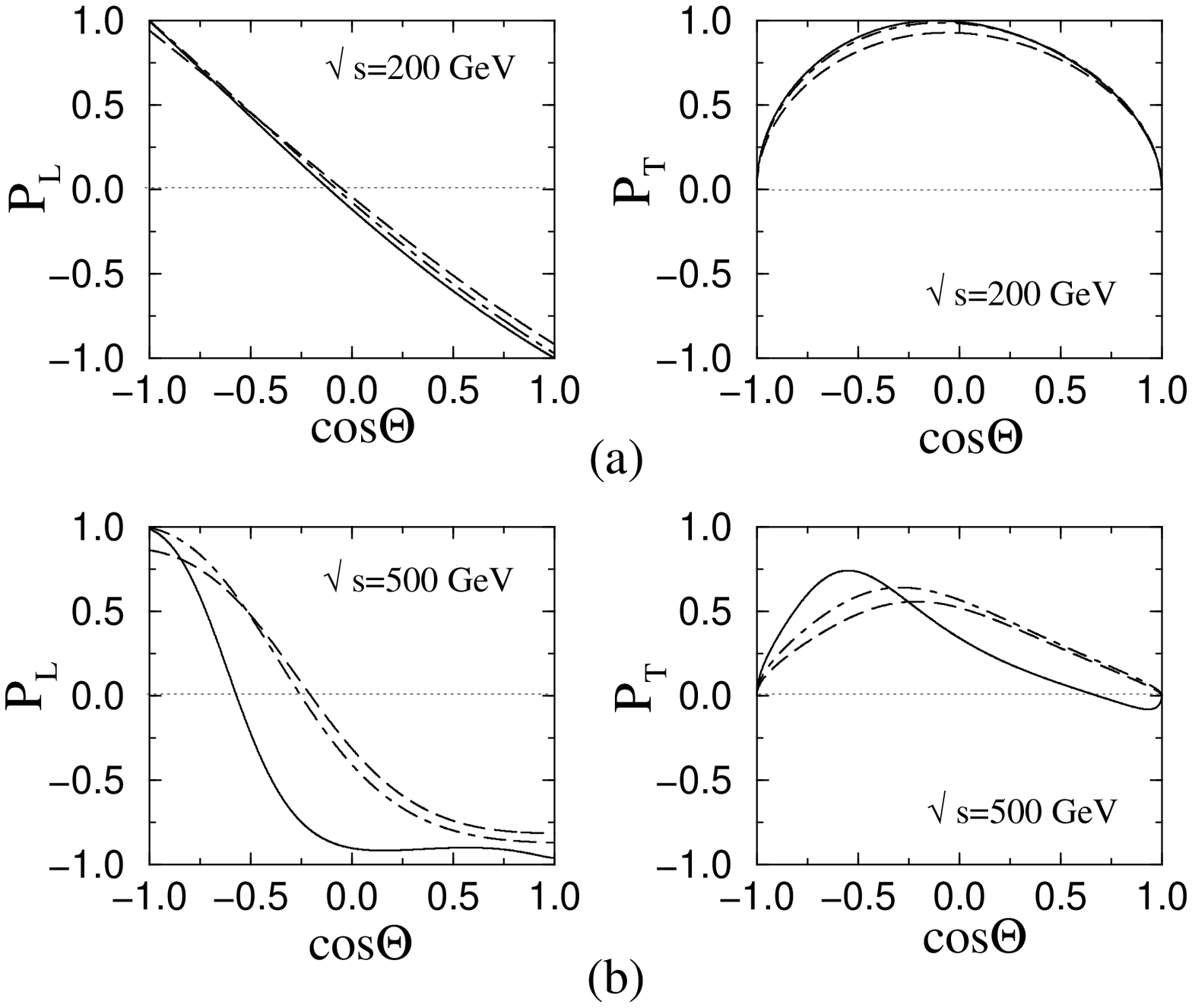,height=12cm}\hss}
\caption{\it The angular dependence of the longitudinal polarization 
             ${\cal P}_L$ and the transverse polarization ${\cal P}_T$ 
             for the same parameters as for the cross section 
             (\ref{eq:parameter}) at  c.m. energies  (a) 200 and (b)
             500 GeV;  solid line  for  the gaugino case, 
              dashed line for the higgsino case, and  dot-dashed line 
             for the mixed case.}
\label{fig:pol}
\end{figure}
\end{center}

\addtocounter{figure}{1}

\newpage
\mbox{ }
\vskip 1cm
\begin{center}
\begin{figure}[htb]
\hbox to\textwidth{\hss\epsfig{file=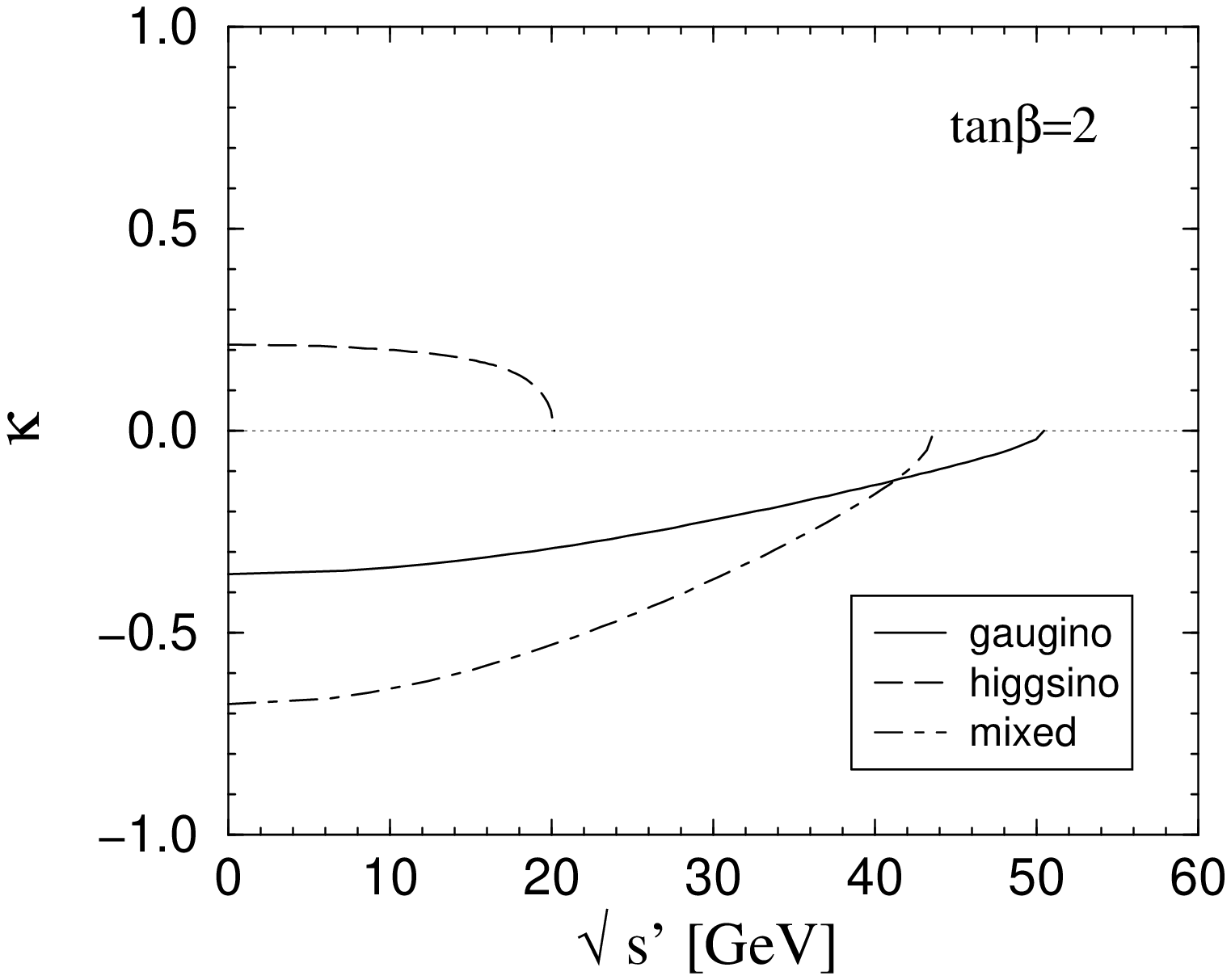,height=7cm}\hss}
\caption{\it The polarization analysis--power $\kappa$ as a function of the
             hadron invariant mass $\protect\sqrt{s'}$ for the same 
             set of parameters as 
             for the cross section (\ref{eq:parameter}). }
\label{fig:kappa}
\end{figure}
\end{center}
%


%
\begin{center}
\begin{figure}[htb]
\hbox to\textwidth{\hss\epsfig{file=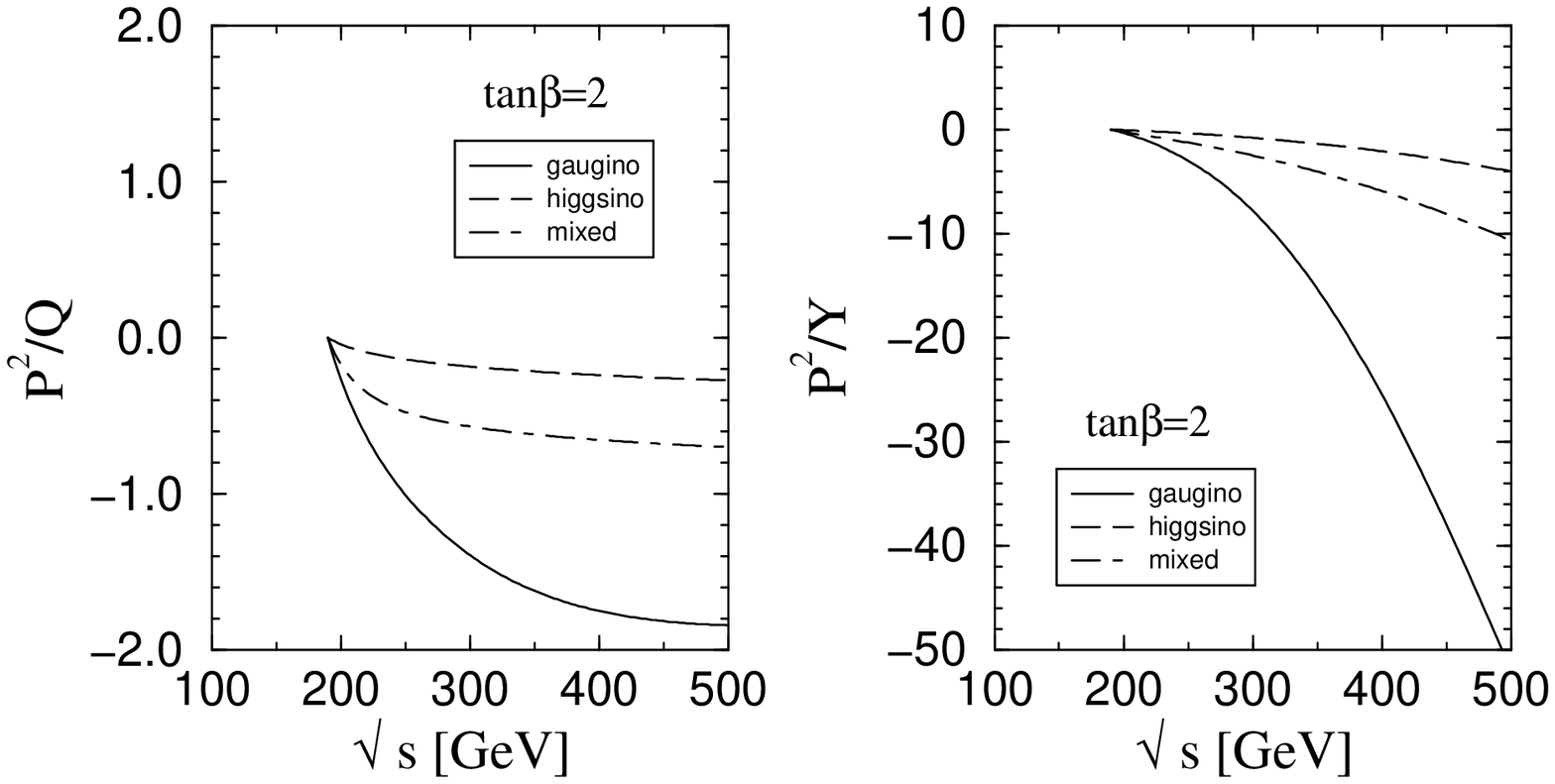,height=6cm}\hss}
\caption{\it The  energy dependence of the ratios ${\cal P}^2/{\cal Q}$ 
             and ${\cal P}^2/{\cal Y}$ for the same 
             set of parameters as for the cross section (\ref{eq:parameter}); 
             solid line  for the gaugino case,  dashed line for  
             the higgsino case, and  dot-dashed line for the mixed case.}
\label{fig:p2qy}
\end{figure}
\end{center}

\newpage
\mbox{ }
\vskip 1cm

\begin{center}
\begin{figure}[htb]
\hbox to\textwidth{\hss\epsfig{file=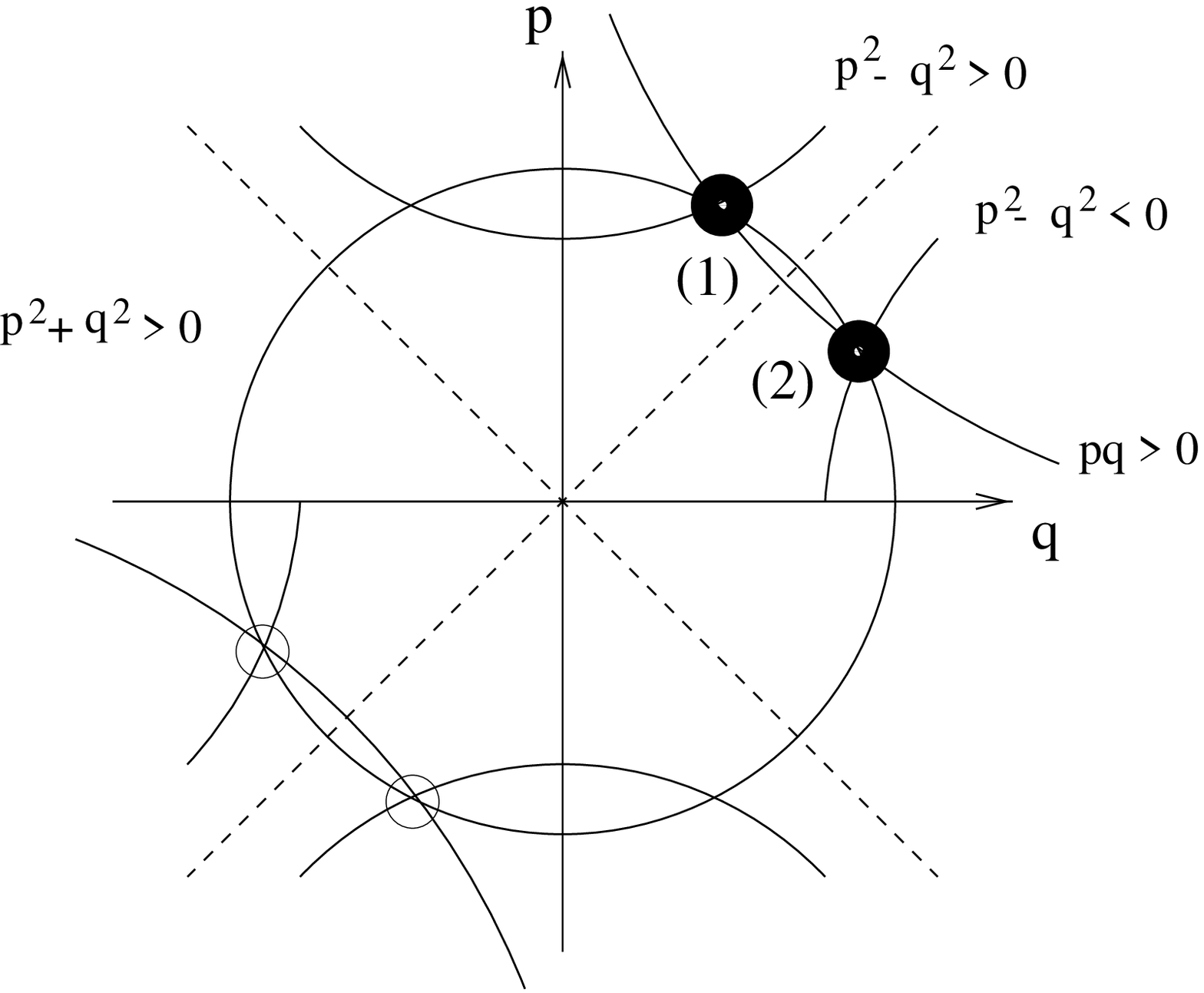,height=6cm}\hss}
\caption{\it Determination of $(p,q)$ from $p^2+q^2$, $pq$ and $p^2-q^2$. 
             The solutions are illustrated for positive values of $pq$.}
\label{fig:pq}
\end{figure}
\end{center}
\vskip -3cm 
\begin{center}
\begin{figure}[htb]
\hbox to\textwidth{\hss\epsfig{file=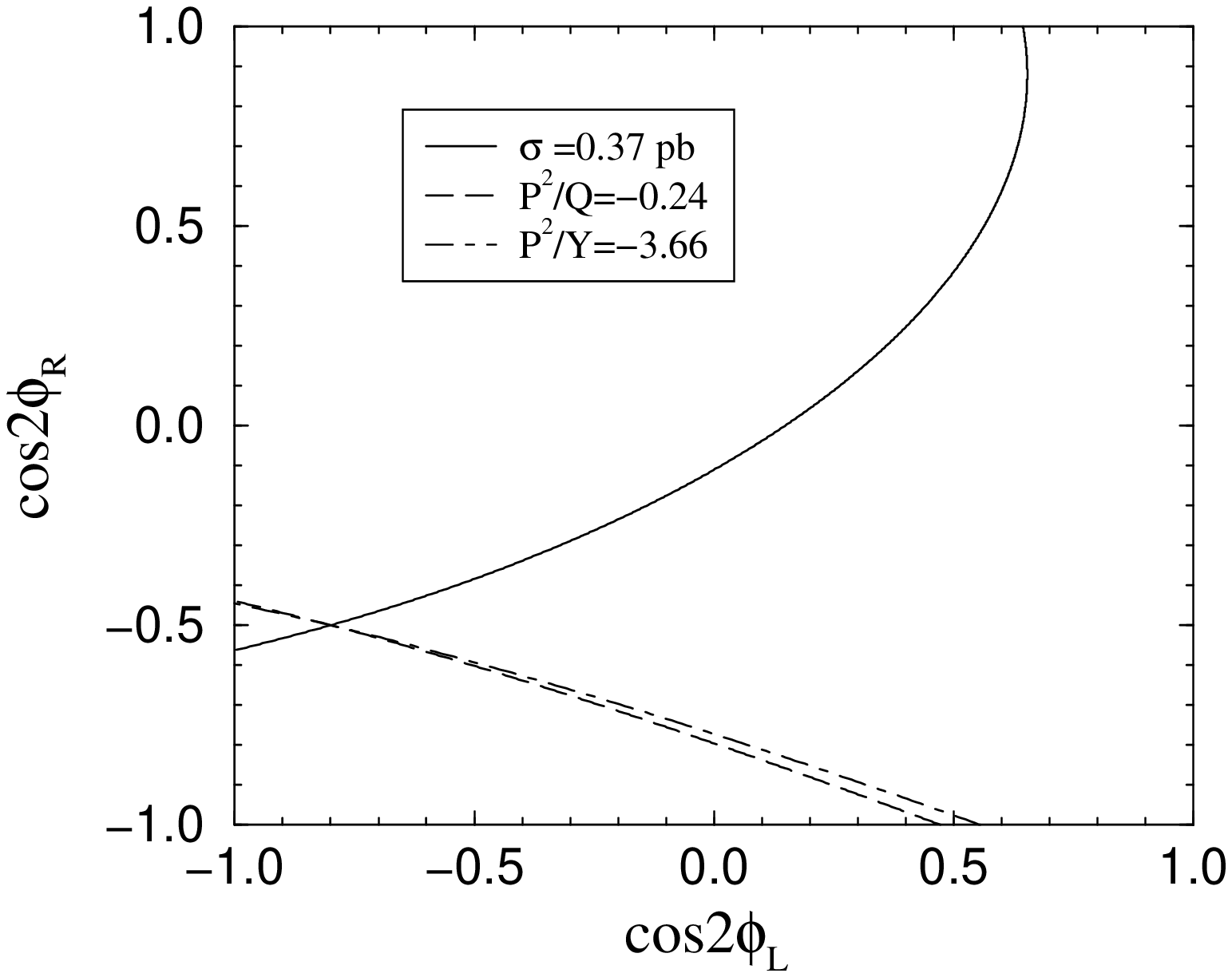,height=7cm}\hss}
\caption{\it Contours for the ``measured values''
             (\ref{eq:measured}) of the total 
             cross section (solid line), ${\cal P}^2/{\cal Q}$ (dashed line), 
             and ${\cal P}^2/{\cal Y}$ (dot-dashed line) 
             for $m_{{\chi}^\pm_1}=95$ 
             GeV [$m_{\tilde{\nu}}=250$ GeV].}
\label{fig:contour}
\end{figure}
\end{center}

\begin{thebibliography}{99}

\bibitem{R1} For reviews of supersymmetry and the Minimal Supersymmetric Standard 
   Model, see H.~Nilles, Phys.~Rep.~{\bf 110} (1984) 1; H.E.~Haber and 
   G.L.~Kane, Phys.~Rep.~{\bf 117} (1985) 75. 

\bibitem{R2} J.~Ellis, J.~Hagelin, D.~Nanopoulos and M.~Srednicki,
   Phys.~Lett.~{\bf 127B} (1983) 233; V.~Barger, R.W.~Robinett, W.Y.~Keung 
   and R.J.N.~Phillips, Phys.~Lett.~{\bf B131} (1983) 372; D.~Dicuss, 
   S.~Nandi, 
   W.~Repko and X.~Tata, Phys.~Rev.~Lett. {\bf 51} (1983) 1030; S.~Dawson, 
   E.~Eichten and C.~Quigg, Phys.~Rev.~{\bf D31} (1985) 1581; A.~Bartl and 
   H.~Fraas and W.~Majerotto, Z.~Phys.~{\bf C30} (1986) 441.  

\bibitem{R3} Proceedings of the Workshop on 
   {\it Physics at LEP II}, Report No. CERN-96-01, eds. G.~Altarelli, 
   T.~Sj\"{o}strand, and F.~Zwirner.

\bibitem{R4} E.~Accomando {\it et al.}, LC CDR Report DESY 97-100 
   (hep-ph/9705442), and Physics~Reports~{\bf 299} (1998) 1.  

\bibitem{R4A} S.Y.~Choi, YUMS 98-3 (hep-ph/9801323); Talk at 
    The First Workshop: Pacific Particles Physics Phenomenology, Seoul 1997.

\bibitem{R6} A.~Leike, Int.~J.~Mod.~Phys.~{\bf A3} (1988) 2895; 
   M.A.~Diaz and S.F.~King, Phys.~Lett.~{\bf B349} (1995) 105; {\bf B373} 
   (1996) 100;  J.L.~Feng and M.J.~Strassler, Phys.~Rev.~{\bf D51}
   (1995) 4461 and {\bf D55} (1997) 1326;  G. Moortgat-Pick and
   H.~Fraas, Report WUE-ITP-97-026 (hep-ph/9708481). 

\bibitem{R6A} G.~Moortgat-Pick, H.~Fraas, A.~Bartl and, and W.~Majerotto,
   WUE-ITP-98-012 (hep-ph/9804306).

\bibitem{R7} L.M.~Sehgal and P.M.~Zerwas, Nucl.~Phys.~{\bf B183} (1981) 
   417.

\bibitem{R8} K.~Hagiwara and D.~Zeppenfeld, Nucl.~Phys.~{\bf B274} (1986) 1.

\bibitem{R9} L.~Michel and A.S.~Wightman, Phys.~Rev.~{\bf 98} (1955) 1190;
   C.~Bouchiat and L.~Michel, Nucl.~Phys.~{\bf 5} (1958) 416;
   S.Y.~Choi, Taeyeon Lee and H.S.~Song, Phys.~Rev.~{\bf D40} (1989) 2477.

\bibitem{R9A} J.H.~K\"uhn, A.~Reiter and P.M.~Zerwas, Nucl.~Phys.~{\bf B272} 
   (1986) 560. 

\bibitem{R11} Y.~Kizukuri and N.~Oshimo, Proceedings of the Workshop on    
   {\it $e^+e^-$ Collisions at 500 GeV: The Physics Potential}, 
   Munich-Annecy-Hamburg 1991/93, DES 92-123A+B, 93-123C, ed. P.~Zerwas;
   T.~Ibrahim and P.~Nath, Phys.~Rev.~{\bf D57} (1998) 478; S.Y.~Choi 
   and M.~Drees, in preparation.

\bibitem{R11A} A.~Datta, M.~Guchait and M.~Drees, Z.~Phys.~{\bf C69} 
   (1996) 347.

\bibitem{R11B} J.H.~K\"uhn and F. Wagner, Nucl.~Phys.~{\bf B236} (1994) 16. 

\bibitem{R12} M.~Brhlik and G.L.~Kane, hep-ph/9803391. 


\end{thebibliography}
\end{document}